\documentclass[12pt]{article}
\usepackage{latexsym}
\usepackage{amsmath}
\usepackage{amssymb}
\usepackage{epsfig,graphics}
\usepackage{graphicx}
\usepackage{booktabs}
\usepackage{multirow}
\usepackage{tikz}
\usepackage{bm}
\usepackage{graphicx}
\usepackage{epsf}
\usepackage{epsfig}
\usepackage{color}
\usepackage{dcolumn}     
\usepackage{amsfonts,amssymb}
\usepackage{dsfont}
\usepackage{slashed}
\usepackage{subfigure}

\begin{document}

\begin{titlepage}

\vspace*{0.6in}
 
\begin{center}
{\large\bf On the spectrum and string tension of \\
$U(1)$ lattice gauge theory in $2+1$ dimensions.}\\
\vspace*{0.75in}
{Andreas Athenodorou$^{a}$ and Michael Teper$^{b}$\\
\vspace*{.25in}
$^{a}$Computation-based Science and Technology Research Center, The Cyprus Institute, 20 Kavafi Str., Nicosia 2121, Cyprus \\
\vspace*{.1in}
$^{b}$Rudolf Peierls Centre for Theoretical Physics, \\
Clarendon Laboratory, University of Oxford,\\
Parks Road, Oxford OX1 3PU, UK}
\end{center}

\vspace*{0.4in}

\begin{center}
{\bf Abstract}
\end{center}

We calculate the low-lying spectra of glueballs and confining flux
tubes in the $U(1)$ lattice gauge theory in $2+1$ dimensions. We see that
up to modest lattice spacing corrections, the glueball states are
consistent with being multiparticle
states composed of non-interacting massive $J^{PC}=0^{--}$ particles.
We observe that the $ag^2 \to 0$ limit is, as expected, unconventional,
and follows the well-known saddle-point analysis of Polyakov to a good
approximation. The spectrum of closed (winding) flux tubes exhibits
the presence of a massive world-sheet excitation whose mass is
consistent with that of the bulk screening mass. These $U(1)$ calculations
are intended to complement existing lattice calculations of the
properties of $SU(N\geq 2)$ and $SO(N\geq 3)$ gauge theories in $D=2+1$.

\vspace*{0.95in}

\leftline{{\it E-mail:} a.athenodorou@cyi.ac.cy, mike.teper@physics.ox.ac.uk}

\end{titlepage}

\setcounter{page}{1}
\newpage
\pagestyle{plain}

\tableofcontents

\section{Introduction}
\label{section_intro}

The compact $U(1)$ lattice gauge theory in $2+1$ dimensions is interesting
because while its dynamics is non-trivial at all non-zero values of the
lattice spacing $a$, it is to a large extent analytically tractable
\cite{Polyakov1,Polyakov2,Polyakov3,BMK77,Mack1,Mack2,Mack3}.
This is in contrast to both the $D=3+1$ $U(1)$ theory, which is essentially
free on the weak coupling side of its strong-weak coupling phase transition,
and also to $D=2+1$ $SU(N)$ gauge theories, which are (largely) analytically
intractable.

While there have been many numerical studies of the `glueball' spectrum of
the $U(1)$ theory in $2+1$ dimensions (for some recent examples see
\cite{Hamer}
or Appendix E of
\cite{MT_SUN}
and references therein), the precision of these older calculations is
inevitably quite poor compared to the precision of recent lattice studies
of $SU(N)$ 
\cite{AAMT_SUN}
and of $SO(N)$
\cite{RLMT_SON}
gauge theories in $D=2+1$,
and it would be useful to have something comparably precise to complement
the latter. That is one of the aims of this paper. Another aim is to
calculate the spectrum of non-contractible closed flux tubes that wind around
one of the periodic spatial directions, so as to learn what we can about the
effective world-sheet action of such flux tubes in the $U(1)$ gauge theory,
thus complementing recent studies
\cite{Caselle_string1,Caselle_string2}
of the ground state of a flux tube between static sources.

In Section~\ref{section_lattice} we describe the compact $U(1)$
lattice gauge theory and also how we calculate the masses of the particles
and the energies of the winding flux tubes. We then move on, in
Section~\ref{section_expectations}, to a summary of what one expects
to find on the basis of earlier analytic work
\cite{Polyakov1,Polyakov2,Polyakov3,BMK77,Mack1,Mack2,Mack3}.
We give some heuristic
descriptions of the screened gas of monopole-instantons and of the
resulting linear confinement, and of the deconfinement transition.
In Section~\ref{section_glueballs} we present our results for the
glueball spectrum, which support the expectation that the theory is a free
theory of (pseudo)scalar particles up to some lattice spacing corrections,
with the mass being just the screening mass of the monopole plasma.
We then calculate the flux tube spectrum, in Section~\ref{section_fluxtubes},
where we find evidence for a massive mode in the world-sheet action of the
flux tube. We summarise our conclusions in Section~\ref{section_conclusions}.

\section{Lattice preliminaries}
\label{section_lattice}

\subsection{action and variables}
\label{subsection_action}

Our U(1) variables are phases, $U_l=\exp \{i\theta_l\}$, assigned to the forward
going links, $l$, of the lattice, and we assign $U^\star_l=\exp\{-i\theta_l\}$ when
we traverse the link backwards. If the link emanates in the direction $\mu$
from the site labelled $n$, an alternative notation is to write $U_l$ as
$U_\mu(n)$ and similarly for $\theta_l$. Under a local gauge transformation
$V(n)=\exp \{i\phi(n)\}$ the phase on the link from the site $n$ to the neighbouring
site in the $\mu$ direction, $n+\hat{\mu}$, transforms as
$U_\mu(n) \to V(n)U_\mu(n)V^\star(n+\hat{\mu})$, just like the path ordered exponential
of the gauge potential along the link in the corresponding continuum gauge theory.
(Although since our theory is Abelian the path ordering is not needed.)
The partition function of the Euclidean lattice gauge theory is
\begin{equation}
  Z 
  =
   \int dU \exp\{-\beta S[U]\}
  =
   \prod_l \int_{-\pi}^{+\pi} d\theta_l \exp\{-\beta \sum_p (1-\cos \theta_p)\},
\label{eqn_Z}
\end{equation}
where $S[U]$ is the action of the lattice field $U$, and $\theta_p$ is the sum of
angles on the links around the plaquette $p$, i.e.
$\theta_p\equiv\theta_{\mu\nu}(n)=\theta_\mu(n)+\theta_\nu(n+\hat{\mu})-\theta_\mu(n+\hat{\nu})-\theta_\nu(n)$,
if $p$ is in the $(\mu,\nu)$ plane emanating from the site $n$, with $\mu<\nu$. Moreover
\begin{equation}
  \beta\stackrel{a\to 0}{\to} \frac{2}{ag^2},
 \label{eqn_beta}
\end{equation}
where $g^2$ is the gauge coupling, which in $D=2+1$ has dimensions of mass, and so
$ag^2$ is the dimensionless coupling on the length scale of the lattice spacing $a$.
We work on periodic $l_x\times l_y\times l_t$ lattices.

\subsection{calculating masses and string tension}
\label{subsection_MK}

We shall give a brief and incomplete summary here. For more details, in particular
of important caveats, see the discussion for the similar $SU(N)$ calculations in
\cite{AAMT_SUN}.

Let ${\cal{C}}(x)$ be a closed and contractible space-like path on the lattice that
begins and ends at the site $x$. Let $\Phi_{\cal{C}}(x)$ be the operator obtained
by taking the product
of the $U_l$ phases on the links that form the boundary of $\cal{C}$. Since U(1) phases
commute, it is obvious that $\Phi_{\cal{C}}(x)$ is gauge invariant. From now on we
assume that we sum our operators over all the spatial sites, so that we have
$p=0$ operators at time $t$, which we label here $\Phi_{\cal{C}}(t)$. We can decompose
the correlator of a generic operator $\Phi$ in terms of the energy eigenstates
$|n\rangle$ and the corresponding eigenstates $E_n$ in the standard way
\begin{equation}
  \langle \Phi^\dagger(t)\Phi(0)\rangle
  =
  \sum_n |\langle n|\Phi|vac \rangle|^2 \exp\{-E_nt\}
  \stackrel{t\to\infty}{=}
  |\langle 0|\Phi|vac \rangle|^2 \exp\{-aE_0n_t\}.
\label{eqn_Mcor}
\end{equation}
Here $|n=0\rangle$ denotes the lightest state such that $\langle 0|\Phi|vac \rangle \neq 0$.
Since in other gauge theories this ground state will normally be a single particle 
we will refer to it as a glueball. We have used $t=an_t$ and since $n_t$ is
our lattice variable we see that what we obtain is the value of $aE_0$, i.e.
the energy in lattice units. By taking linear combinations of rotated
operators and adding/subtracting reflections we can form operators that have a
spacific spin $J$ and parity $P$. Moreover the real part is $C=+$ and the imaginary part
is $C=-$. So to obtain the lightest mass with specific $J^{PC}$ quantum numbers
we simply take correlators of operators with those quantum numbers. Excited
states will be obtained through a variational extension of eqn(\ref{eqn_Mcor})
\cite{AAMT_SUN}.

There are various important caveats including the assignment of $J$ on a lattice, and 
the nature of finite volume corrections, and we refer the reader to
\cite{AAMT_SUN}
for a more complete discussion of these. Here we mention some practical considerations.
Since our Monte Carlo calculation of the correlator in eqn(\ref{eqn_Mcor}) has
statistical errors that are roughly independent of $t$, while the exact correlator
decreases exponentially in $t$, it is clear that if the asymptotic exponential
is to be visible we need an operator with a large enough overlap onto the ground
state that this asymptotic exponential already  dominates the correlator at small $t$.
This can usually be achieved by using operators that are iteratively `blocked'
\cite{MT_block,MT_SUN,AAMT_SUN}.
To evaluate the credibility of the identification of an asymptotic exponential
it is useful to define an effective energy
\begin{equation}
  C(n_t)
  =
  \frac{\langle \Phi^\dagger(n_t)\Phi(0)\rangle}{\langle \Phi^\dagger(n_t-1)\Phi(0)\rangle}
  =
  \exp\{-aE_{eff}(n_t)\}.
\label{eqn_Eeff}
\end{equation}
If $aE_{eff}(n_t)$ is consistent with being constant for $n_t \geq n_0$ then
it is consistent with being described by a single asymptotic exponential
for  $n_t \geq n_0$ and we obtain an estimate of the energy $aE_0$ from the
corresponding single exponential fit. Since the statistical error on  $aE_{eff}(n_t)$
grows exponentially in $n_t$ such evidence for an asymptotic exponential
is not necessarily convincing: one needs several consecutive values of $n_t$ 
in the fit of  $C(n_t)$ for which  $aE_{eff}(n_t)$ has small errors. This
can be undermined even for small $aE_0$ if the overlap is small so that
$n_0$ is large. But even if the overlap is good, but  $aE_0$ is large,
then the evidence can be weak. That is to say, our calculations of masses
and energies will be most reliable for the states that are simultaneously
light and have a large overlap onto our basis of operators.
We shall show explicit examples later in the paper.

It is also important to note that because of the positivity properties
of the expansion in eqn(\ref{eqn_Mcor}) we know that the exact
value of $aE_{eff}(n_t)$ must be monotonically decreasing with $n_t$.
And since the typical error in identifying the effective mass plateau
is to begin it at too small a value of $n_t$, the effect of this error
is usually to overestimate the energy. So in these cases we should treat
mass estimates as upper bounds on the true mass.

To calculate the confining string tension we calculate the energy 
of a confining flux tube that winds around, say, the $x$ direction
of our finite periodic lattice. If this length is $l$ we denote this
energy by $E_f(l)$. The calculation is the same as described above for
the glueballs except that the operators are built on curves $\cal{C}$
that are closed but non-contractible: they wind once around the $x$-circle.
For large $l$ we can obtain the string tension  $\sigma$ from the fact that
$E_f(l)\to \sigma l$ as $l\to\infty$. However since  $E_f(l)$ becomes large
at large $l$, and so difficult to calculate reliably, we will in practice
calculate  $\sigma$ from intermediate values of $l$ where higher order
corrections to the linear behaviour may be important. These higher
order corrections are accurately given by the formula
\begin{equation}
  E_f(l)
  =
  \sigma l \left(1-\frac{\pi}{3\sigma l^2}\right)^{\frac{1}{2}}
\label{eqn_NG}
\end{equation}
which encodes all the known universal corrections
\cite{Aharony1,Aharony2,Aharony3,Aharony4}
and which is known to work well in $SU(N)$
\cite{AAMT_string1,AAMT_string2,AAMT_string3}
and $SO(N)$ gauge theories
\cite{RLMT_SON}
down to small values of $l$. We shall show below that it works equally
well for $U(1)$.

\section{U(1) expectations}
\label{section_expectations}

Naively one expects the continuum $U(1)$ theory to be a free field theory
of massless photons. That is to say, if one takes the limit
$\beta=2/ag^2\to\infty$ and expresses quantities in units of $g^2$
then what one obtains is a free theory of massless particles.
Of course if one places charged static sources a distance
$r$ apart into the theory then one obtains a non-trivial Coulomb potential
which is logarithmically confining, $V(r) = \{g^2/\pi\}\log r$ in $D=2+1$. 
If the theory were non-Abelian, then perturbation theory would produce
an additional series of higher order corrections in increasing powers of
$g^2r$ (the dimensionless coupling on the scale $r$) but this of course
could only be trusted when $g^2r$ is small. In fact for large $r$
we know that in any gauge theory the potential $V(r)$ cannot grow
faster than $\propto r$ 
\cite{Seiler}.
(Heuristically, in a confining flux tube a faster growth would imply a
diverging energy density and hence a breakdown of the vacuum.) In our
Abelian $U(1)$ case perturbation theory simplifies and we only have
the logarithmic Coulomb piece up to lattice spacing corrections,
although this may cease to hold in a finite volume
\cite{Copeland}.
Of course, as we shall 
summarise below, in the $U(1)$ lattice gauge theory the vacuum contains a
dilute gas of instantons which drive the screening of the $U(1)$ fields, so
that the perturbative contributions will be exponentially suppressed
at large enough $r$ for any non-zero lattice spacing.

As remarked above, at finite $a$ the compact $U(1)$ lattice gauge theory
possesses non-trivial physics which is driven by the instantons of the theory
\cite{Polyakov1,Polyakov2,Polyakov3,BMK77}.
In Euclidean $D=2+1$ the field of this instanton is just like the field
of a magnetic monopole in 3 space dimensions. So we shall follow
convention and refer to the instantons as monopoles or monopole-instantons,
and we shall often refer to their fields as `magnetic', although
these fields point out in both timelike and spacelike directions.
These are singular Dirac monopoles with a diverging ultraviolet action
and this makes possible reliable saddle-point calculations
\cite{Polyakov1,Polyakov2,Polyakov3}.
One finds that the magnetic fields are screened with a
screening mass $m_D$ that sets the screening length
\cite{Polyakov1,Polyakov2,Polyakov3}.
This is the mass gap, the lightest particle in the theory, and it possesses
$J^{PC}=0^{--}$ quantum numbers. Thus the vacuum contains a screened plasma
of monopole-instantons and, in addition, one can show that this generates
a linearly confining potential betwen static charged sources, with a
string tension $\sigma$. One finds, using $\beta=2/ag^2$,
\cite{Polyakov1,Polyakov2,Polyakov3,Mack1,Mack2,Mack3}
\begin{equation}
  a^2m^2_D \stackrel{\beta\to\infty}{=} c\beta \exp\{-\tilde{c}\beta\}
\label{eqn_mD}
\end{equation}
and
\begin{equation}
  a^2\sigma \stackrel{\beta\to\infty}{=} c^\prime ag^2 am_D.
\label{eqn_K}
\end{equation}
Here the (semi)classical values of the various constants are
\begin{equation}
  \tilde{c}=0.2527\pi^2,\qquad c=\sqrt{8\pi^2},\qquad c^\prime=\frac{8}{\sqrt{2\pi^2}}
\label{eqn_KmDconst}
\end{equation}
where $\tilde{c}$ is directly related to the action of an isolated monopole
on the lattice. We see that both $m_D/g^2$ and $\surd\sigma/g^2$
vanish as $\beta=2/ag^2 \to \infty$. Thus we have a trivial
massless free field theory if we take the continuum limit keeping
$g^2$ as our fixed scale. However since
$g^2$ is not directly physical one might instead choose to take
the continuum limit keeping $m_D$ fixed. In this case we see
from the above that $\sigma/m^2_D \to \infty$ as $a\to 0$,
so any putative glueballs formed of closed loops of flux will
become infinitely massive and will decouple from the theory. 
Thus we are left with a theory of free massive scalar particles
in this way of defining the continuum limit. The above are heuristic
arguments, but can be made much more precise by saddle point calculations
\cite{Polyakov1,Polyakov2,Polyakov3}
and the main conclusions can be made rigorous
\cite{Mack1,Mack2,Mack3}.

Because the monopoles play a fundamental role in the physics of the
$U(1)$ theory we describe them in more detail in
Section~\ref{subsection_monopoles} and explain why we have screening.
(For a more extensive and pedagogical analysis see for example
\cite{Copeland,Wensley1,Wensley2}.)
At the qualitative level it is very simple to see how such monopoles
produce linear confinement, and we provide the argument in 
Section~\ref{subsection_confinement}. We then discuss the
deconfining transition at finite temperature, together with some
numerical illustrations, in Section~\ref{subsection_deconfinement}.
Finally we make some observations on the consequences of
taking different scales to define our continuum limit.
It is the fact that the monopoles are ultraviolet objects that
makes possible simple heuristic arguments that capture the
essential dynamics. At the same time the fact that monopoles
consist of near-maximal field fluctuations on the scale of $a$ means
that our computer simulations inevitably encounter a severe
form of critical slowing down which limits how far in $\beta$
we can easily go.

\subsection{monopoles (instantons) and screening}
\label{subsection_monopoles}

To motivate how one identifies the presence of a monopole-instanton in
a $U(1)$ lattice gauge field it is useful 
to think of the lattice field as being embedded in a continuous
$U(1)$ gauge field. It will be convenient, because the $D=2+1$ space-time
is  Euclidean, to use a notation that is appropriate to 3 space dimensions.
For example, our `magnetic' flux will not be the (pseudo)scalar of 2 space
dimensions but rather the vector of 3 space dimensions.

We recall that the variable on each link $l$ is
the phase $\exp\{i\theta_l\} \equiv \exp\{i\theta_\mu(n)\}$ where $l$ is the
link emanating in the positive $\mu$ direction from the lattice site $n$
and $\theta_l \in (-\pi,+\pi]$.
In the continuum limit, with fields smooth on the scale of $a$,
$\theta_\mu(n)$ becomes simply the gauge potential in lattice units.
To be more precise, given how $\exp\{i\theta_l\}$ transforms under a gauge
transformation, it corresponds to the exponential
of the integral of gauge potential along the link $l$, i.e.
\begin{equation}
\theta_\mu(x) \sim \int_x^{x+a\hat{\mu}}ds A_\mu(s).
\label{eqn_thetaA}
\end{equation}
So if we sum the angles on the link matrices around a plaquette $p$ 
(conjugated when they are backward going) then we obtain the total magnetic
flux $B_p$ through any surface which is bounded by that plaquette's
boundary and on which the fields are not singular
\begin{equation}
 \theta_p =  \sum_{l\in\partial p} \theta_l = \int_{\partial p} ds A_l(s) = B_p,
\label{eqn_thetapB}
\end{equation}
where $A_l$ is the component along the link $l$.
Note that the flux has a direction and reverses sign if we
reverse the order in which we go around the plaquette.
(Clearly $\theta_l \to -\theta_l$ under this operation.)
Consider then a lattice cube $c$. It is clear that the sum of
the angles around the bounding plaquettes of $c$ must be zero
because for each link matrix on a plaquette there is also its
conjugate from a neighbouring plaquette, i.e.
\begin{equation}
 \sum_{p\in\partial c} \theta_p = 0.
\label{eqn_thetaC}
\end{equation}
Indeed this is true for any closed surface, not just a cube.
Thus if there is no singularity of the gauge potential
on any of the plaquettes then 
\begin{equation}
  \sum_{p\in\partial c} B_p = 0,
\label{eqn_thetaBC}
\end{equation}
i.e the net flux out of the cube is zero
and we have no magnetically charged object within the cube.
(Here $p$ is an oriented plaquette so that $B_p$ is a
flux emanating out from the centre of the cube.)
Now suppose we have large fluctuations on the plaquettes
around a cube. In particular let us suppose the angles sum to $\pi/3$
on each of 5 of the 6 plaquettes. Then by eqn(\ref{eqn_thetaC})
we know that on the 6'th plaquette we have $B_p=-5\pi/3=\pi/3 - 2\pi$
if there are no singularities on this plaquette.
So now the question we must ask is, do we in fact have
a singularity on this plaquette? If we have a minimal
singularity then encircling the singularity provides the $-2\pi$.
and the actual flux will be just $\pi/3$.
That is to say, we have a minimally charges Dirac monopole 
within the cube with a Dirac string piercing the 6'th plaquette.
Since the energy (action) density is $\propto B^2$, it is clearly
much more likely that there is a Dirac string piercing the
sixth plaquette, and hence a magnetic flux $\pi/3$, rather than
no string and hence the much larger  magnetic flux of $-5\pi/3$
through the plaquette. This simple argument 
essentially transposes to the lattice the elegant argument by
Dirac in the continuum theory
\cite{Dirac}.
This way of identifying monopoles is the standard lattice algorithm
\cite{DeGrand_Toussaint}:
one decomposes each sum of plaquette angles into
\begin{equation}
  \theta_p = 2\pi n_p + \tilde{\theta}_p
  \quad ; \qquad \tilde{\theta}_p \in (-\pi,+\pi)
\label{eqn_thetaM}
\end{equation}
and the monopole charge in the cube is then the sum of the
integers $n_p$ on the oriented plaquettes bounding the cube.
If we now consider such a monopole on the lattice then the
same argument can be applied to any larger surface enclosing
the monopole. If we have a monopole that is a long way
from other (anti-)monopoles then a small smooth deformation
of the fields cannot remove it (although it may move it
to a neighbouring cube). Thus an isolated monopole in
an infinite volume is at a minimum of the action (which
may in general include zero-modes).
That is to say these are instantons which can form the
basis for a saddle-point calculation. Since a monopole's
presence in a cube is tied to a Dirac string leaving the cube
through one of its faces (or a net number of Dirac strings
leaving the cube) and since each face is shared with a
neighbouring cube, it is clear that in a finite periodic volume
a monopole must be partnered by an anti-monopole so that
the net magnetic charge is zero and the net magnetic flux
out of the space-time volume is also zero.

We should mention that much of the analytic work has been performed
using the Villain lattice action
\cite{Villain}
which is in the same universality class as our plaquette action.
In addition to the angles on the links it contains explicit integer
valued variables living on the plaquettes which when summed over
the faces of a cube can be shown to provide a direct definition
of the monopole charge within that cube. It is natural to ask
how closely these monopoles are related to those that are
extracted using eqn(\ref{eqn_thetaM}). The numerical answer
is that outside the region near $\beta\sim 0$ of very strong
coupling, the monopole gases obtained by these two definitions
differ almost always by no more than a number of small dipoles
which have no impact on the long distance physics of the theory
\cite{ZSMT_Villain}.

Since the monopoles are pointlike and have an action that is
ultraviolet large, they would appear to be ideal candidates
for being treated as a dilute gas of instantons. However
if we consider such a dilute gas then we immediately see a
problem: the interaction energy/action of the monopole with
the dilute gas diverges. This is because the potential energy with
a single (anti-)monopole $\propto \pm 1/r$. Within a shell
a distance $r$ from the monopole we have a number of monopoles
and anti-monopoles that is proportional to the volume of the shell,
i.e. $n_M(r),n_{\bar{M}}(r) \propto r^2$. This is a dilute gas so
the modulus of the net magnetic charge is
$|n_M(r)-n_{\bar{M}}(r)| \propto \sqrt{r^2} \sim R$ and the contribution
to our original monopole's energy is
$\propto \pm r\times 1/r \sim \pm r^0$. Integrate this over $r$
with random signs and one has a divergence. To control these
divergences the gas needs to have some order rather than being
completely random: this will obvously decrease the entropy
but may render the energy well defined. The answer is well known
from the physics of a plasma of charged particles
in three spatial dimensions: the particles
of the gas order themselves in such a way as to induce screening.
At large distances the field of our monopole is screened so that
its field falls exponentially with $r$ and the interaction
energy is finite. The inverse screening length defines a screening
mass, $m_D$, which more careful calculations show to be given by
eqn(\ref{eqn_mD}). This is essentially a screened photon
and has $J^{PC}=0^{--}$ quantum numbers.

\subsection{monopoles and confinement}
\label{subsection_confinement}

Once we know that the monopoles organise themselves into a screened
gas it is straightforward to predict the qualitative impact on the
potential $V(r)$ between static charges a distance $r$ apart.
Consider a rectangular Wilson loop of extent $r \times \tau$ in
the spatial and temporal directions respectively. Then, as is well
known, if we integrate the gauge potential around the Wilson loop and
exponentiate it and take the expectation value and take $\tau\to\infty$,
we will obtain the potential between the two static charges.
This is true for general gauge theories and here we can write it as
\begin{equation}
  \langle \exp\{i\int_{r \times \tau}Ads\}\rangle
  =\langle \exp\{iB(r,\tau)\}\rangle
  \stackrel{r,\tau \to \infty}{=} c(r)\exp\{-V(r)\tau\},
\label{eqn_WL}
\end{equation}
where we use the fact that integrating the potential around the loop
gives us the magnetic flux passing through the rectangular surface
spanned by the loop, up to an additive integer multiple of $2\pi$
coming from Dirac strings which clearly does not affect the value of
$\exp\{iB(r,\tau)\}$.

To calculate the contribution of the monopole gas to
$B(r,\tau)$ we make the (inessential) approximation that
the flux of any  monopole within a slab extending a distance of
$l_D=1/m_D$ either side of the Wilson loop is not screened at all
and the gas therein is random, while monopoles further away are
completely screened. Thus $B(r,\tau)$ only receives contributions
from monopoles within the space-time slab. Assuming a symmetric
flux from each monopole and given that $r$ and $\tau$ are arbitrarily
large, half of the flux, i.e. $\pi$ units, will pass through the
rectangle. (We neglect the vanishing fraction of monopoles near the
boundary.)
Given that
this is a random dilute gas, we can use the Poisson distribution
for the probability of having a particular number of (anti)monopoles.
In this approximation we can factorise the contributions of
monopoles and antimonopoles, and those from the lower and upper slabs
since the magnetic flux is additive. Moreover the contribution
of any (anti-)monopole to $B(r,\tau)$ is $\pm \pi$ and when
exponentiated in eqn(\ref{eqn_WL}) this just gives a factor
of $\exp\{\pm i\pi\} = -1$. We can place each monopole
anywhere in the volume $l_D r\tau$ of the slab, 
and the probability of finding one in a unit volume is
$c(\beta) \propto \exp\{-\beta S_M\}$ where $S_M$ is the action
on the lattice of an isolated monopole. Thus we have the estimate
\begin{equation}
  \langle \exp\{iB(r,\tau)\}\rangle
  =
  \left(\sum_n \frac{(c(\beta)l_D r\tau)^n}{n!}e^{-\bar{n}}(-1)^n\right)^4
  =
  e^{-8\bar{n}}
  =
  e^{-8c(\beta)l_D r\tau},
\label{eqn_WLb}
\end{equation}
where the average number of monopoles is $\bar{n}=c(\beta)l_D r\tau$
in such a Poisson distribution, and the power of 4 comes from
multiplying the contribution of the slabs on both sides and doing all
this for monopoles and antimonopoles separately. Comparing to
eqn(\ref{eqn_WL}) we see that the monopoles produce a linearly
confining potential between the charges
\begin{equation}
  V(r) = 8c(\beta)l_D r = \frac{8c(\beta)}{m_D} r = \sigma r
\label{eqn_V_M}
\end{equation}
with a string tension $\sigma$. Given the various approximations we made
in this argument, we should not expect our expression for $\sigma$  to match 
the exact  expression for $\sigma$ in eqn(\ref{eqn_K}). However since
$c(\beta)$ in eqn(\ref{eqn_V_M}) is proportional to 
$\exp\{-\tilde{c}\beta\}\propto m_D^2$ in  eqn(\ref{eqn_mD}), we see
that in  eqn(\ref{eqn_V_M}) we have $\sigma \propto m_D$ just as in  
eqn(\ref{eqn_K}). In any case the approximations do not affect the
qualitative argument and the above calculation captures the reason why a 
screened monopole gas produces a linearly confining potential between
static charges.

\subsection{deconfinement}
\label{subsection_deconfinement}

As we have seen above the monopole-instantons in the $U(1)$ lattice gauge
theory in an infinite Euclidean space-time volume generate linear confinement.
This corresponds to a temperature $T=0$. At some finite $T=T_c$ we can
expect the theory to deconfine. If the correlation length diverges
at $T_c$ then one expects
\cite{SY_Tc}
the transition to be in the same universality class as the spin model
in one lower dimension that encodes the centre symmetry of the gauge theory.
Here the centre is $U(1)$ and we are in $D=2+1$, so the spin model is the
$XY$ model in $D=2$. So the  relevant phase transition is the
Kosterlitz-Thouless transition which is of infinite order
\cite{KT}.
Such a transition is challenging to study numerically although some
lattice studies of the the $U(1)$ transition and its critical
exponents do exist
\cite{U1_Tc_MC1,U1_Tc_MC2,U1_Tc_MC3}.
We do not intend to study the transition in this paper but we do need
to know enough about it to be confident that our calculations in this
paper are indeed in the low-$T$ confining phase.
We will begin by showing how we expect $T_c$ to scale and then
we will display some typical numerical signals of the transition.

We start with a well-known general argument and then proceed to a
more specific dynamical argument. Consider a flux tube attached to sources a
large distance $l$ apart at temperature $T$. The contribution to the
partition function $Z(T)$ is 
\begin{equation}
\delta Z(T)
\sim
N(l) \exp\left\{ -\frac{E(l)}{T} \right\},
\label{eqn_dZTa}
\end{equation}
where $N(l)$ is the number of flux tubes between the sources, i.e.
the `entropy' factor, and $E(l)$ is the energy of the flux tube.
(In this heuristic argument we make the simplifying approximation
that the flux tubes are all of length $l$.) For large $l$ we
have $E(l)\simeq \sigma l$ where $\sigma$ is the string tension, and
one can easily show that the dominant behaviour of  $N(l)$ is
exponential in $l$ with the scale set by the flux tube width $l_0$
i.e.
\begin{equation}
\delta Z(T)
\propto
\exp\left\{\frac{cl}{l_0}\right\} \exp\left\{-\frac{\sigma l}{T}\right\}
=
\exp\left\{\left(\frac{c}{l_0}-\frac{\sigma }{T}\right)l\right\},
\label{eqn_dZTb}
\end{equation}
where the dimensionless constant $c=O(1)$ depends on details, and we ignore
sub-leading power factors in $N(l)$ since they do not affect the argument.
It is clear from eqn(\ref{eqn_dZTb}) that as we increase $T$ at
some value $T=Tc$ the entropy factor will overwhelm the energy factor and
$\delta Z(T)$ will diverge as $l\to\infty$. So for $T>Tc$ the vacuum
contains a condensate of arbitrarily long flux tubes and so it costs
nothing to increase $l$, i.e. the theory is deconfining. From
eqn(\ref{eqn_dZTb}) we see that this occurs at 
\begin{equation}
T = T_c = \frac{\sigma l_0}{c}
=  \frac{\tilde{c} \sigma}{c m_D},
\label{eqn_Tc}
\end{equation}
where we assume that the width of the flux tube is $l_o=\tilde{c}/m_D$
with $\tilde{c}=O(1)$ since the screening mass, $m_D$, 
is the fundamental dimensionful dynamical quantity in the theory.
Now from eqns(\ref{eqn_mD},\ref{eqn_K})  we see that this means that
\begin{equation}
  aT_c \propto 1/\beta
  \Longrightarrow \frac{T_c}{g^2}\stackrel{a\to 0} {=} \mathrm{constant}.
\label{eqn_Tcg}
\end{equation}
This is interesting and somewhat counterintuitive: whereas the
other nontrivial dynamical quantities such as $m_D$ and $\sigma$ have
the unusual behaviour $\propto \exp\{-c/ag^2\}$ as $a\to 0$
(arising from the ultraviolet nature of the instanton dynamics)
the deconfining temperature has a conventional scaling continuum limit.

We now complement the above general argument with a derivation of $T_c$
that relies more closely on the $U(1)$ dynamics, and which will therefore
allow us to obtain more details of the transition. We saw earlier that
linear confinement can be attributed to the flux from the monopole-instantons
that are within a screening length $l_s = 1/am_D$ of the Wilson loop.
The calculations rely on a symmetric radial flux from the monopoles but
this is only exactly true in a Euclidean space-time volume that is
infinite in all directions, and hence is at $T=0$. For $T>0$ the
timelike extent is $L_t=1/T$. Since the boundary is periodic the
magnetic flux cannot pass through it and will be directed in a purely
spatial direction once it is far enough from the monopole. So if we are
a distance $r \gg 1/T$ from the monopole the $2\pi$ of flux will be
equally distributed over a cylindrical surface of area
$A=2\pi r L_t =2\pi r/T $ so that the flux density is $B=2\pi/A = T/r$.
To obtain the total action $E_M(R)$ of the monopole field
within a distance $R \gg 1/T$ we integrate the action density $B^2$ over
the radius $r$, the angle $\theta$ and Euclidean time $\tau$ to obtain,
approximately,
\begin{equation}
  E_M(R)
  \propto \int^{1/T}_{0}d\tau \int^R_a rdr \int^{2\pi}_0 d\theta B^2(r)
\propto 2\pi T \log R.
\label{eqn_EM}
\end{equation}
So as the spatial volume increases to infinity, the action of a monopole
diverges logarithmically with the spatial size. This is of course precisely
the behaviour of vortices in the $D=2$ $XY$ model. This is no surprise:
for $R\gg L_t=1/T$ the monopole fields are in a thin slab pointing radially
outwards and thus directly mimicking a $D=2$ vortex with radially pointing
spins. And the argument for a phase transition is the same as in the
XY model. Consider a monopole in a space-time volume $R^2L_t$. Its
contribution to the partition function will be 
\begin{eqnarray}
\delta Z(R,T)
\propto
R^2L_t \exp\{-\beta  E_M(R)\}
& \propto &
\exp\{2\log R\} \exp\{-2\pi \beta T \log R \} \nonumber \\
& \propto &
\exp\{2(1-\pi T \beta) \log R \}
\label{eqn_dZRT}
\end{eqnarray}
since the monopole can be placed anywhere in the volume $R^2L_t$ (its entropy)
and we have used eqn(\ref{eqn_EM}) and $L_t=1/T$ (and we drop factors that
can be neglected). We see that because the divergence of the action with
$R$ is only logarithmic it can be overcome by the trivial entropy factor
that arises from translating the position of the monopole; so
as $R\to\infty$ the weight of the monopole vanishes
for $T > 1/\pi\beta$ and diverges for $T < 1/\pi\beta$, indicating a phase
transition at $T=T_c=1/\pi\beta$. All this can be trivially extended to a
dilute gas of monopoles and so we expect a phase transition at
\begin{equation}
  T_c = \frac{c}{\beta} 
\label{eqn_Tc2}
\end{equation}
for some constant $c$ which is just $1/\pi$ in the simple calculation above.
For $T< T_c$ we have a gas (condensate) of monopoles which will produce a
linearly confining force between charges, while for  $T > T_c$ the entropy
no longer wins out over the action and any monopole cannot exist in isolation
but must be bound to a neighbouring (anti-)monopole in a dipole, i.e. the
monopoles are `confined' and the argument for the linear confinement of
charges breaks down.

All this strongly suggests that the transition should be in the same
universality class as that of the XY model, i.e. of infinite order.
Some earlier lattice calculations have attempted to calculate the
critical exponents of this transition and these appear to be
consistent with that expectation
\cite{U1_Tc_MC1,U1_Tc_MC2,U1_Tc_MC3}.
We note that the above argument only works if $L_t=1/T < 1/m_D$ where
$m_D$ is the screening mass. If $L_t> 1/m_D$, so that the monopole
fields are screened away before they reach the time boundary, then
the monopole will not be affected by the temporal periodicity.
However, since $1/T_c \propto \beta$ while $1/m_D \propto \exp\{+c\beta\}$,
we will always satisfy $1/T < 1/m_D$ when $T$ is near $T_c$ as long
as $\beta$ is not too small.

\begin{figure}[htb]
\begin	{center}
\leavevmode
\input	{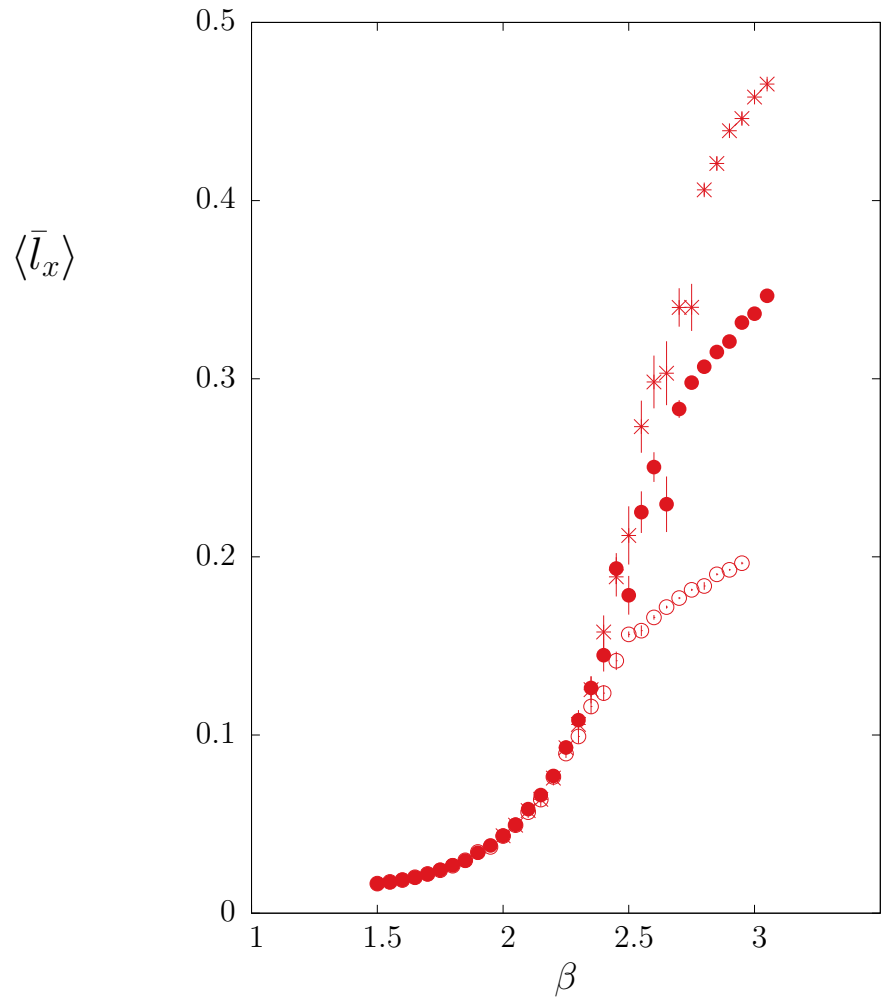}
\end	{center}
\caption{Values of $\langle \bar{l}_x \rangle$ on $8\times 96\times 96$, $\star$,
  $8\times 64\times 64$, $\bullet$, and $8\times 32\times 32$, $\circ$, lattices,
  rescaled to common values at smallest $\beta$.}
\label{fig-Lx-t8b}
\end{figure}

\begin{figure}[htb]
\begin	{center}
\leavevmode
\input	{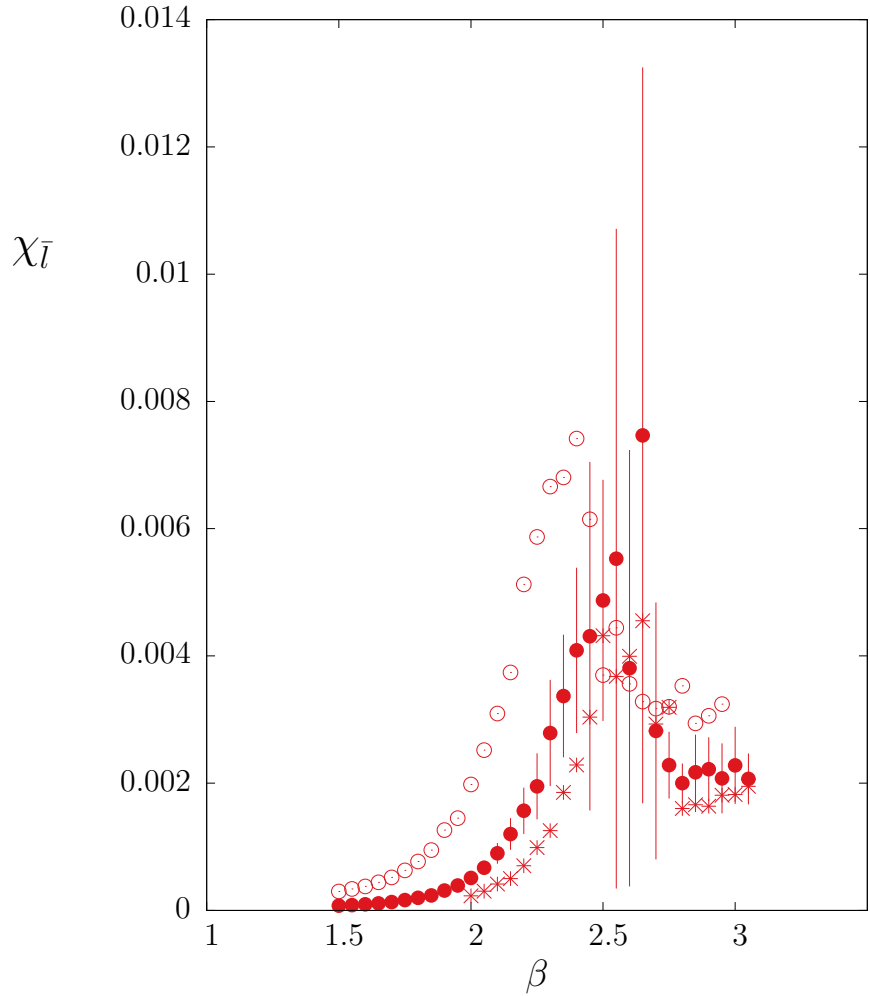}
\end	{center}
\caption{The (vacuum-subtracted) susceptibility of $|\bar{l}_x|$ on
  $8\times 32^2$ ($\circ$), $8\times 64^2$ ($\bullet$) and $8\times 96^2$ ($\star$)
  lattices. For clarity only errors on the $l=64$ points are shown.}
\label{fig-Lsusc-t8}
\end{figure}

Although we do not intend to study deconfinement in this paper --
all our calculations are intended to be in the $T \sim 0$ confined
phase -- we wish to reassure ourselves that our values of $L_t$
do indeed satisfy $T=1/L_t \ll T_c$. We have therefore performed
calculations on $L^2L_t$ lattices with $L_t=8$ and $L=32,64,96$
to identify the approximate location in $\beta$ of the deconfining
transition. We do this in the conventional way by first taking the
Polyakov loop, which is defined as the product of link variables along
a circle that winds once around the temporal circle. Call it  $l(x,y)$
where $(x,y)$ is the spatial position in the time-slice where we
start the loop. We then average $l(x,y)$ over  $x,y$
and take the modulus of this average, before averaging over all fields:
i.e. we use as our `order parameter' for the transition
\begin{equation}
  \langle |\bar{l}| \rangle
  =  \langle |\frac{1}{N_s}\sum_{x,y} l(x,y)| \rangle,
\label{eqn_modL}
\end{equation}
where $N_s$ is the number of lattice sites in a time-slice.
The reason for this somewhat ugly modification of the Polyakov loop
expectation value is the same as in $SU(N)$ gauge
theories: for $T>T_c$ the centre symmetry is spontaneously broken
in an infinite volume, but in a finite volume (such as ours) the system
will tunnel through to the various vacua so that in a sufficiently long
calculation we will find $\langle {l} \rangle = 0$ for all $T$. On the
other hand $\langle |\bar{l}| \rangle$ will be the same in all these vacua
(the typical lattice field will be in one vacuum) so it suffers
no cancellations and  should be `large' in the deconfined phase.
This is at a price:  $\langle |\bar{l}| \rangle$ will also be non-zero
in the confined phase, but this value will be small and will vanish
as $L\to\infty$. So while this variable is a little ugly
theoretically, it can serve the purpose of identifying the
crossover between phases on a finite volume.
So we plot  $\langle |\bar{l}| \rangle$ on our $L^28$ lattices
in Fig.~\ref{fig-Lx-t8b}. For clarity we have renormalised
the data for the different $L$ so that they have the same value at
$\beta=2.0$, which clearly lies away from the transition region.
We see a very clear cross-over from small to large values with the
steepest rate of change somewhere around $\beta_c \sim 2.5$, although
$\beta_c$ increases slightly with increasing $L$, something which
would have been more apparent if we had not renormalised the data.
If we had not renormalised then we would have seen that
$\langle |\bar{l}| \rangle$
decreases with increasing $L$. So this has the characteristics
of the expected deconfining transition.  It is useful and conventional
to complement this analysis with the susceptibility of the same
quantity. This we display in   Fig.~\ref{fig-Lsusc-t8}
for our three spatial volumes: $L=64$. This confirms that the
transition lies in the region $\beta_c \sim 2.5$. These calculations
are not an attempt to add to earlier work
\cite{U1_Tc_MC1,U1_Tc_MC2,U1_Tc_MC3}
but simply to tell us how large a lattice we need to use to 
be well within the confining phase in our range of $\beta$.
Given that $L_t^c=1/T_c \propto \beta$ to leading order, we infer
that we will need our lattices to have $L,L_t >> 8\beta/2.5$ to
satisfy this requirement. 

We remarked that $T_c/g^2$ is non-zero as $\beta\to\infty$. However as is
apparent from the above, all the physical differences between the two phases
will disappear $\propto \exp\{-c\beta\}$ as $\beta\to\infty$, so despite its
conventional scaling, the phase transition will effectively disappear if we
approach the continuum limit in the conventional fashion by taking
$g^2$ as our physical scale. If on the other hand we take $m_D$ as our
physical scale, then in this continuum limit $T_c/m_D \to \infty$ and the
deconfinement transition disappears from the theory.

\subsection{continuum limit(s)}
\label{subsection_continuum}

As we remarked above, a formal treatment of the $U(1)$ theory tells us
that the continuum theory is a free field theory of massive particles
which, moreover, is linearly confining at any non-zero lattice spacing
\cite{Mack1,Mack2,Mack3}.
In the  $U(1)$ lattice theory there are several possible length scales
that one might consider keeping fixed as one approaches the continuum
limit, $a\to 0$. These are the inverse coupling, $l_g = 1/g^2$,
the inverse string tension, $l_\sigma = 1/\surd\sigma$, the inverse
screening mass, $l_D = 1/m_D$, and the average separation between
the monopole-instantons, $l_M$. Expressions relating the first three
length scales are given in eqns(\ref{eqn_mD},\ref{eqn_K}), so we
see that $l_D \sim g^2l^2_\sigma$.  The scale
$l_M$ gives us the average number of monopoles in a space-time volume
$\bar{n}(V) = V/l_M^3$. So from eqn(\ref{eqn_V_M}) we deduce that
\begin{equation}
  l^3_M \sim l^2_\sigma l_D.
\label{eqn_lM}
\end{equation}
It is interesting to see what kind of `continuum limit' one might
obtain in keeping each of these length scales fixed. In asking this
question we need to express our lengths and energies in units of
the fixed physical length and energy scale (with the latter being
the inverse of the former).

\noindent {(1) {\it $l_g$ fixed as $a \to 0$}}. In this limit the
average separation between the monopole-instantons grows
$\propto \exp\{+c/ag^2\}$ where $c>0$ so on any fixed length scale
$r$ such that $r/l_g < \infty$, there are no monopoles at all in the
$a\to 0$ continuum limit and there is no linear confinement. That
is to say if we introduce charged sources a distance $r$ apart
such that $r/l_g < \infty$ then the potential energy is simply
the logarithmic Coulomb one. We also have $m_D=0$ in this limit so
that the theory is a free field theory of massless `photons'.
Of course in this free continuum field theory the coupling $g^2$ does
not provide a scale, since it can be absorbed into the fields,
whereas in the lattice theory it cannot be simply absorbed away
because the fields are integrated over a compact range so the
coupling would reappear in the range of integration.

\noindent {(2) {\it $l_\sigma$ fixed as $a \to 0$}} In this limit
physical distances are some finite multiple of $l_\sigma$ and
physical energies are some finite multiple of
$E_\sigma=1/l_\sigma=\surd\sigma$. From eqn(\ref{eqn_K}) we see that
\begin{equation}
  \frac{m_D}{E_\sigma} = \frac{\surd\sigma}{g^2}
  \stackrel{\beta\to\infty}{\longrightarrow} 0,
\label{eqn_lK}
\end{equation}
i.e. we have a theory of massless particles. If we insert charged
sources then we can consider separately the linear potential,
$V_\sigma = \sigma r$, and the Coulomb potential, $V_c\sim g^2\log r$.
If we change the distance from $r_1=\hat{r}_1 l_\sigma$ to 
$r_2=\hat{r}_2 l_\sigma$ the change in the linear potential in
physical units is $\delta V_\sigma/E_\sigma =  (\hat{r}_2 - \hat{r}_1)$
which is finite. However this is deceptive: the linear potential can
only appear on scales $r\gg 1/m_D$ and since
$m_D/E_\sigma \stackrel{\beta\to\infty}{=} 0$ we have in fact no linear
potential at any finite physical distance in this continuum limit.
(Although it will be there at finite $\beta$ at large enough $r$.)
What we have at finite $r$ is just the Coulomb potential, which
is not screened since $m_D=0$ in this continuum limit. The change
in that potential in going from $r_1=\hat{r}_1 l_\sigma$ to 
$r_2=\hat{r}_2 l_\sigma$ is, in physical units,
\begin{equation}
  \frac{\delta V_c}{E_\sigma}
  =
  \frac{g^2 \log\{\hat{r}_2 /\hat{r}_1\}}{\pi \surd\sigma}
  \stackrel{\beta\to\infty}{\longrightarrow} \infty .
\label{eqn_dVc}
\end{equation}
Thus we cannot couple charged particles to the theory in this limit.

\noindent {(3) {\it $l_D$ fixed as $a \to 0$}}
In this case the screening mass in physical units is finite by
definition since $m_D/E_D =1$ where $E_D=1/l_D=m_D$.
So the theory is one of massive particles. If we now couple
charged sources a distance $r=\hat{r}l_D$ apart then we will have a
linear potential $V_\sigma \simeq \sigma r$ as long as $r\gg l_D$,
and a Coulomb potential for $r\ll l_D$, which becomes screened
at larger $r$. The change in the linear potential when we increase
$r$ from $r_1$ to $r_2$ is 
\begin{equation}
  \frac{\delta V_\sigma}{E_D}
  =
  \frac{\sigma (\hat{r}_2 l_D - \hat{r}_1 l_D)}{E_D}
  =
  \frac{\sigma (\hat{r}_2 - \hat{r}_1)}{m^2_D}
 =
  \frac{g^2}{m_D} (\hat{r}_2 - \hat{r}_1)
  \stackrel{\beta\to\infty}{\longrightarrow} \infty ,
\label{eqn_dVK}
\end{equation}
where we have used eqn(\ref{eqn_K}). A similar argument shows that
the change in the Coulomb potential is
$\delta V_c/E_D = (g^2/\pi m_D) \log\{\hat{r}_2/\hat{r}_1\}$
which also diverges as $\beta\to\infty$. So in this case
as well we cannot couple charged particles to the theory.

\noindent {(4) {\it $l_M$ fixed as $a \to 0$}} If we use $l_M$  as
our fixed unit of length then, by definition, our fields will
contain a finite non-zero density of monopoles. The corresponding
energy scale is $E_M=1/l_M$ and one can easily see, using
eqns(\ref{eqn_lM},\ref{eqn_mD},\ref{eqn_K}), that
\begin{equation}
  \frac{m_D}{E_M} = \{g^2l_D\}^{-\frac{1}{3}}
  \stackrel{\beta\to \infty}{\longrightarrow} 0
\label{eqn_mlM}
\end{equation}
so this continuum limit has massless particles. If we now
consider the linear confining potential $\sim \sigma r$
between two sources a distance $r$ apart, and if we ask by
how much this potential changes between $r_1=\hat{r}_1l_M$
and  $r_2=\hat{r}_2l_M$ then we find that this change, in
units of our energy scale $E_M$ is divergent:
\begin{equation}
  \frac{\delta V_\sigma}{E_M}
  =
  \frac{\sigma(\hat{r}_1 - \hat{r}_2)l_M}{E_M}
  =
  \frac{l^2_M}{l^2_\sigma}(\hat{r}_1 - \hat{r}_2)
  \propto \{g l_M\}^{\frac{1}{2}}
  \stackrel{\beta\to \infty}{\to} \infty.
\label{eqn_dVM}
\end{equation}
Similar manipulations show that the same is true for
the Coulomb potential. All this means that we cannot add charged
sources (or particles) to the theory if this is to constitute a sensible
continuum limit. \\

What we infer from the above is that in none of these candidate continuum
limits do we see a linearly confining potential. In some the string
tension diverges in physical units, while in others the transition from
Coulomb to linear only occurs at infinite separation. (Of course
on the lattice at finite $\beta$ there are no divergences.) A corollary
is that in none of these limits do we have non-perturbative glueballs
composed of contractible closed loops of flux. Either such states
are ultraviolet, or the `flux tube' width, $\sim 1/m_D$, is infinite.
So the only particle is the screened `photon'.

\section{Glueballs and string tension}
\label{section_glueballs}

As remarked above, the expectation is that in the continuum limit
we have a free theory of massive scalar particles. So when we calculate
the mass spectrum we expect the lightest particle to be a `massive photon' 
in the $J^{PC}=0^{--}$ sector with the mass $m_D$ given in eqn(\ref{eqn_mD}).
All other states will be states composed of two or more of these
particles which will be free up to finite lattice spacing corrections.
Thus we expect the lightest $0^{++}$ state to consist of two of these
photons with zero relative momentum, and the first excited $0^{--}$ state 
to consist of three of these photons with zero relative momentum, i.e.
\begin{equation}
  m_{0^{--}}^{gs} = m_D \quad ; \quad  m_{0^{++}}^{gs} = 2m_D \quad ; \quad  m_{0^{--}}^{ex1} = 3m_D.
  \label{eqn_massesG}
\end{equation}
Other states will have photons with momenta that are non-zero, and these
need not be free momenta due to phase shifts induced by interactions
due to lattice spacing corrections. Moreover they will typically have
small gaps with other excitations (on a large volume) so their analysis
is more complicated and goes beyond the scope of this paper.
So we shall focus on the ground and first excited $0^{--}$ states and
the ground $0^{++}$ state, and see whether we can confirm the expectation
in eqn(\ref{eqn_massesG}), and also that in  eqn(\ref{eqn_mD}), at least
as far as the functional dependence on $\beta$ is concerned.

The tension of the confining string is the other quantity we calculate
in this section. While one is used to continuum limits in which
dimensionless mass ratios such as $\surd\sigma/m_D$ would be non-zero
and finite, the prediction of eqns(\ref{eqn_mD},\ref{eqn_K}) is that
this ratio will diverge exponentially in $\beta$, and that it is
the quantity $a^2\sigma\beta/am_D$ that has a finite limit. We will
try to find some numerical evidence for this.

We remark that the glueball masses we will obtain in this Section are
consistent with those obtained in
\cite{Hamer,Caselle_string1,Caselle_string2}
but with much smaller errors. Our string tensions on the other hand,
while also being consistent with the values in
\cite{Caselle_string1,Caselle_string2},
are  considerably less precise than the latter, because of the very different
techniques employed. However our calculations are useful and complementary
both because they are for closed flux tubes rather than the open flux tubes of
\cite{Caselle_string1,Caselle_string2},
and also, more importantly, because we are able to calculate the energies
of excited flux tubes, as we will do in Section~\ref{section_fluxtubes}.

\subsection{glueball masses}
\label{subsection_glueballM}

\begin{figure}[htb]
\begin	{center}
\input	{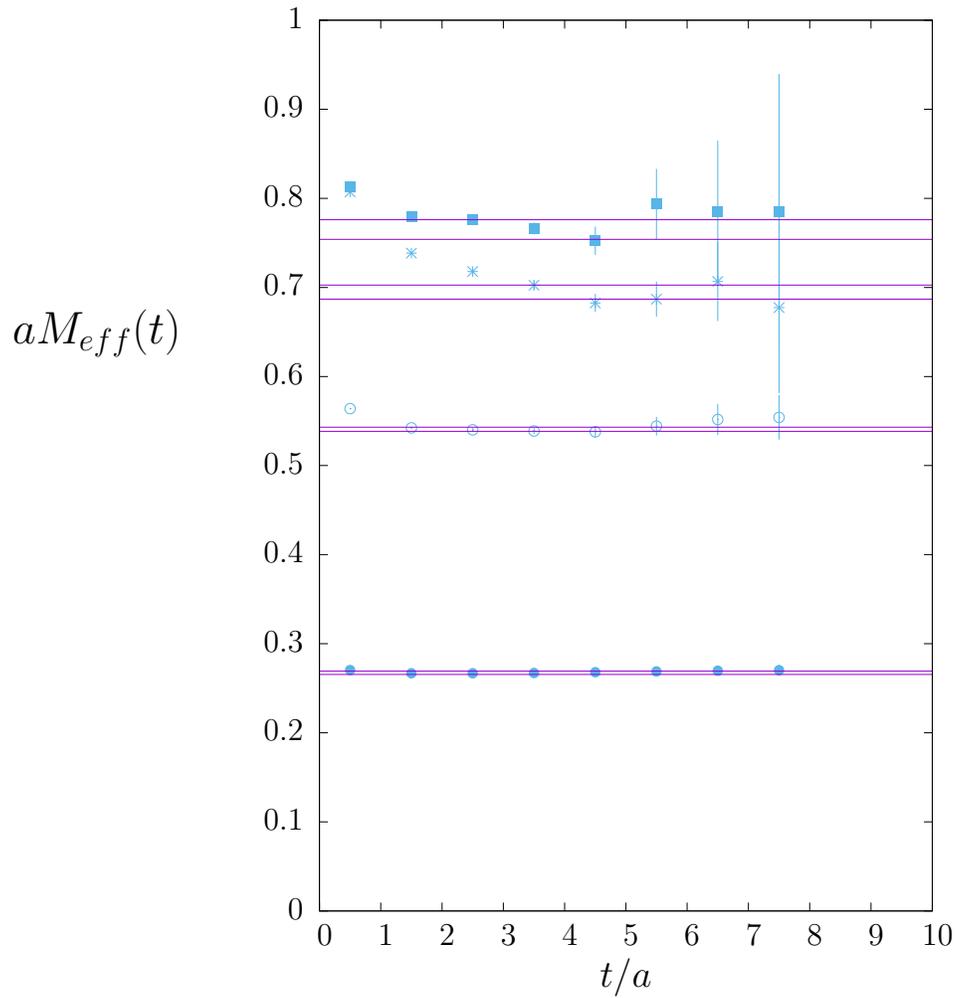}
\end	{center}
\caption{Effective masses for the $0^{--}$ ground state ($\bullet$), the
  $0^{++}$ ground state ($\circ$), the $0^{--}$ first excited state ($\blacksquare$)
  and the  $2^{-+}$ ground state ($\star$). All on a $50^236$ lattice at $\beta=2.2$.
  Pairs of lines are our $\pm 1\sigma$ estimates of the corresponding masses.}
\label{fig_MeffGb2.2} 
\end{figure}

We will calculate the lightest few glueball masses at two couplings,
$\beta=2.2$ and $\beta=2.3$, and on a range of volumes. This will allow us
to check for both finite lattice spacing and finite volume corrections.
But before doing so we need to assess how reliably we can calculate these
masses from our correlation functions. To illustrate this we show in
Fig.\ref{fig_MeffGb2.2} the effective masses extracted from the correlators
of our best variationally selected operators. This is taken on one of our
larger lattices at $\beta=2.2$. We see that the lightest $0^{--}$ and $0^{++}$ 
have unambiguous effective mass plateaux and excellent overlaps.
The first excited  $0^{--}$ has a reasonably convincing plateau and a
good overlap. The $2^{-+}$ has a significantly worse overlap and hence
a less convincing plateau. Other ground states, apart from the $2^{++}$
which is very similar to the $2^{-+}$, are heavier and the plateaux
are harder to identify, so we not use that data here. As stated earlier,
our main focus in the paper will be on the lightest $0^{--}$ and $0^{++}$
masses and on the mass of the first excited $0^{--}$, all of which are
reasonably well determined.

We also show in Fig.\ref{fig_MeffGb2.2} the $\pm 1\sigma$ error bands
on the mass estimates obtained from the correlators. These are purely
statistical errors and we see, for example, that the estimate of the
error is smaller for the $2^{-+}$ than for
the excited $0^{--}$. On the other hand it is clear that any systematic
error associated with identifying the effective mass plateau is
larger for the $2^{-+}$. The caveat from this is that for heavier
particles the quoted errors can be severely underestimated.

Our estimates of the masses of the lightest $0^{--}$ and $0^{++}$ and the
first excited $0^{--}$ are listed in Tables~\ref{table_massV_b2.2}
and ~\ref{table_massV_b2.3} for $\beta=2.2$ and $2.3$ respectively.
We see that the mass gap has $0^{--}$ quantum numbers, just as expected.
While the mass gap exhibits no obvious finite volume corrections,
this is not the case for the $0^{++}$ and the excited $0^{--}$. We plot
the $0^{++}$ and excited $0^{--}$ masses in
Figs.\ref{fig_VGb2.2},\ref{fig_VGb2.3} for $\beta=2.2,2.3$
respectively. We also show horizontal lines corresponding to once, twice
and thrice the large-volume mass gap. We observe that at both values
of $\beta$ we have $M_{0^{++}}^{gs} = 2M_{0^{--}}^{gs}$ within our quite
small errors, once the volume is large enough, roughly $lM_{0^{--}}\geq 9$.
At $\beta=2.3$ we also see that $M_{0^{--}}^{ex} = 3 M_{0^{--}}^{gs}$ within
errors on large volumes, although at $\beta=2.2$ there is a slight
undershoot, which is presumably due to the interactions arising from
higher order lattice corrections which will be larger at smaller $\beta$.

\begin{figure}[htb]
\begin	{center}
\input	{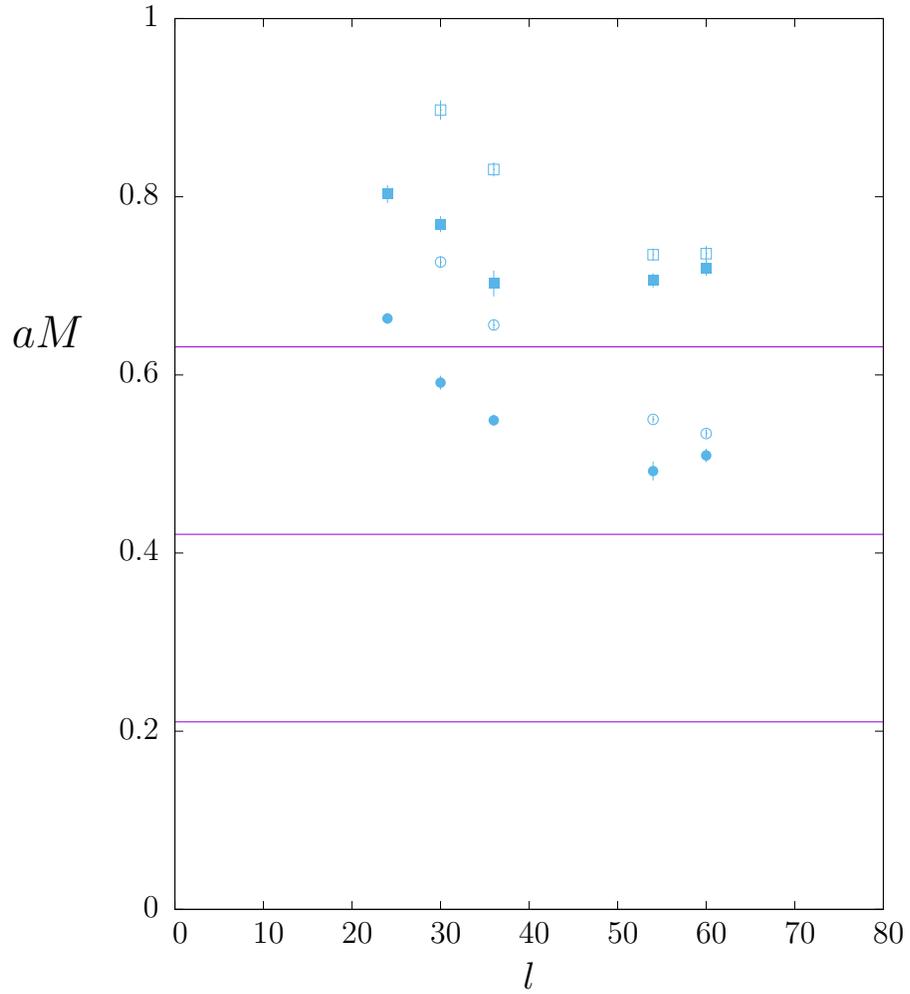}
\end	{center}
\caption{Ground state glueballs: $2^{++}$ ($\bullet$), $2^{-+}$ ($\circ$)
  $2^{--}$ ($\blacksquare$) and $2^{-+}$ ($\square$) on spatial volumes $V=l^2$
  at $\beta=2.3$. Lines are once, twice and thrice the large volume $0^{--}$
  ground state mass. }
\label{fig_VG2b2.3}
\end{figure}

Although we have seen in Fig.\ref{fig_MeffGb2.2} that the $J=2$ mass
is less reliable (the systematic error is surely larger than the
quoted statistical error) it is interesting to see what we find.
In Fig.~\ref{fig_VG2b2.3} we plot the $2^{\pm +}$ masses obtained
on various volumes at $\beta=2.3$. We also show the  $2^{\pm -}$ masses;
these are heavier and so a priori less reliable, but it so happens that
their overlap is somewhat better and this partly compensates.
The expectation in a free field theory is that the $2^{\pm +}$ will
be composed of two $0^{--}$ particles, just like the $0^{++}$, but that
these will have non-zero relative momenta so as to project onto $J=2$.
As we decrease $l$ the allowed momenta grow larger and so will the
energy of the state (although lattice interactions will shift the actual
momenta, e.g. by phase shifts) while as $l\to\infty$ we should see
$M_{2^{\pm +}} \to 2M_{0^{--}}$ albeit slowly. For the $2^{\pm -}$ things
are similar except that we need three $0^{--}$ particles so as to arrive
at a net $C=-$, and so we expect $M_{2^{\pm -}} \to 3M_{0^{--}}$ as $l\to\infty$.
Within the large errors, what we see in Fig.~\ref{fig_VG2b2.3} roughly
supports this scenario.

In summary what we see is consistent with the expectation that the $0^{++}$
ground state and the excited  $0^{--}$ are composed of 2 and 3
non-interacting  $0^{--}$ particles respectively up to lattice spacing
corrections. As the volume decreases these  $0^{--}$ particles will
necessarily be closer and we would expect the effect of the lattice
corrections on the energies of these states to be magnified, just as we
observe. We shall indeed see below that the observed finite volume
corrections do appear to disappear as we increase $\beta$.

\subsection{string tension}
\label{subsection_stringK}

\begin{figure}[htb]
\begin	{center}
\input	{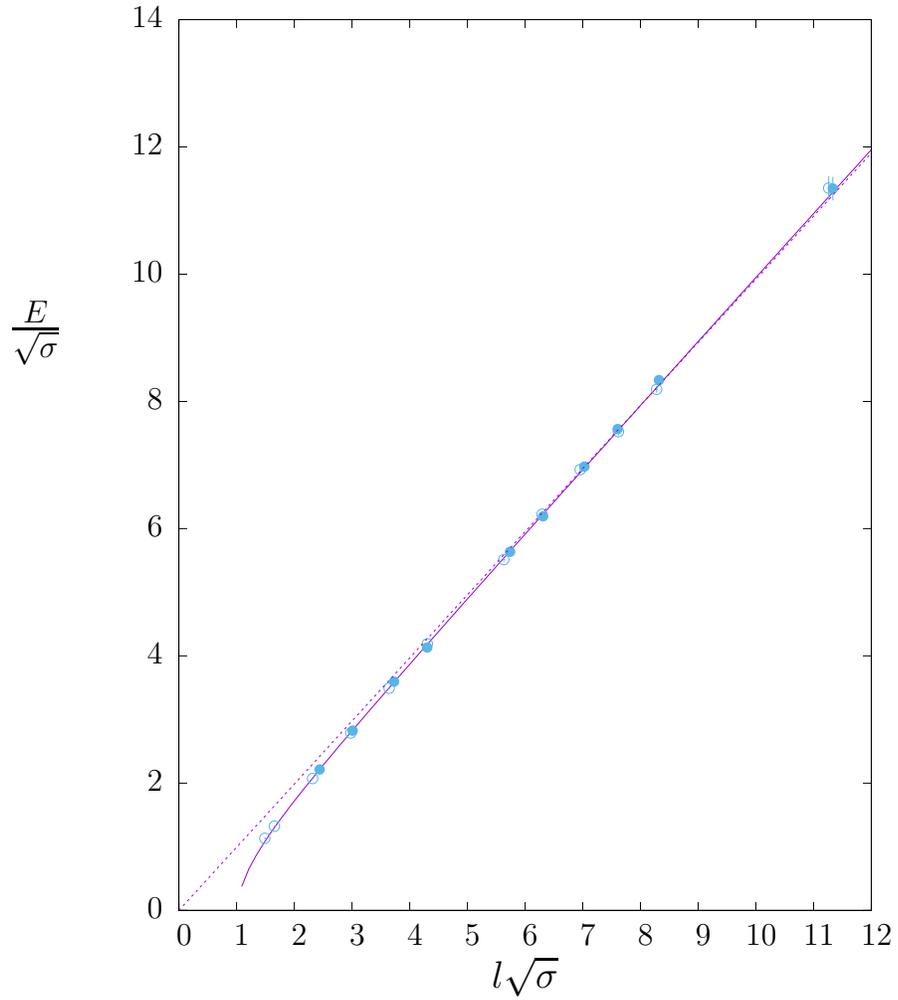}
\end	{center}
\caption{Ground state flux tube energy versus length $l$, at $\beta=2.2$,
  $\circ$, and at $\beta=2.3$, $\bullet$.  Solid line is rescaled form of
   eqn(\ref{eqn_NG}) and dashed line is asymptotic $E=\sigma l$ piece.}
\label{fig_Ek1q0_gs}
\end{figure}

On our periodic lattice we calculate the string tension $\sigma$ from
the energy of a confining flux tube that winds once around a spatial
circle. This energy, $E_f(l)$, will depend on the spatial size $l$
and, in the well studied case of $SU(N)$ gauge theories
\cite{AAMT_string1,AAMT_string2,AAMT_string3},
can be reliably related to $\sigma$ by the expression in eqn(\ref{eqn_NG}),
if $l$ is not very small. The fact that  eqn(\ref{eqn_NG}) encodes all
the universal corrections of the effective string action describing
a winding flux tube
\cite{Aharony1,Aharony2,Aharony3,Aharony4}
suggests it will be equally applicable in the $U(1)$ case, but clearly
this needs to be checked. Accordingly we have performed
calculations of $aE_f(l)$ for various $l$ at $\beta=2.2,2.3$
as listed in Tables~\ref{table_stringb2.2},\ref{table_stringb2.3}.
The ground state of the flux tube has positive parity and appears in the
Tables as $E^{P=+}_{gs}$. Fitting this data with the expression in
eqn(\ref{eqn_NG}), we find that we can get the following good fits
over most of our range of $l$:
\begin{equation}
  a^2\sigma = 0.027423(41)  \quad ; \quad \frac{l}{a}\geq 14\quad
  \frac{\chi^2}{n_{df}}=0.68 \quad \beta=2.2
  \label{eqn_sigmab2.2}
\end{equation}
and 
\begin{equation}
  a^2\sigma = 0.020581(32)  \quad ; \quad \frac{l}{a}\geq 17\quad
  \frac{\chi^2}{n_{df}}=0.81 \quad \beta=2.3.
  \label{eqn_sigmab2.3}
\end{equation}
We can conveniently plot our results by scaling both the ground state
energy $E_f$ and the flux tube length $l$ by the measured value of
$\surd\sigma$ (which is essentially determined by large $l$) so that
all quantities plotted become
dimensionless: this allows us to place the $\beta=2.2$ and  $\beta=2.3$
results on the same plot. This we do in Fig.~\ref{fig_Ek1q0_gs} where we
compare our values to the appropriately rescaled version of eqn(\ref{eqn_NG}).
We see that the data is well described by  eqn(\ref{eqn_NG}) and to
emphasise this point we also plot the asymptotic $E_f=\sigma l$ fit,
showing that the corrections to the linear piece that are encoded in
eqn(\ref{eqn_NG}) are indeed important over a significant range of small $l$. 

\begin{figure}[htb]
\begin	{center}
\input	{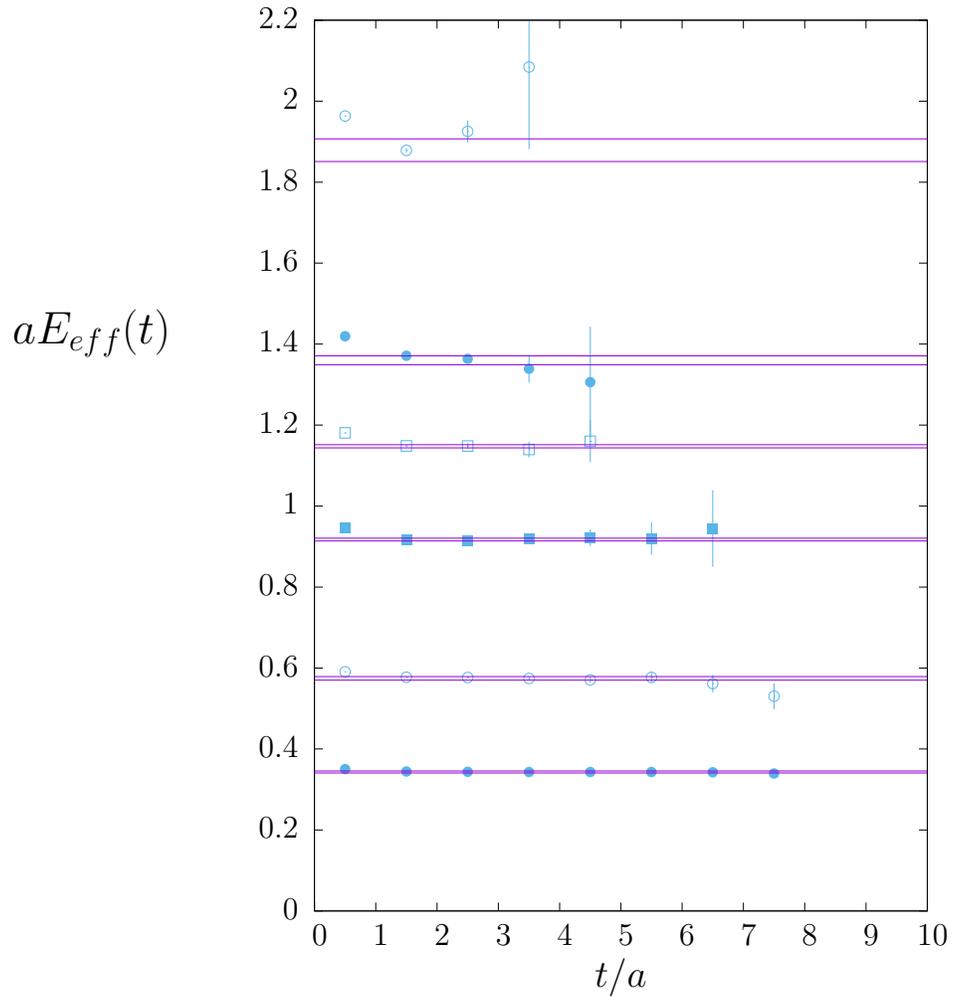}
\end	{center}
\caption{Effective energies of the ground states of winding flux tubes
  with lengths $l=14,22,34,42,50,68$ in increasing order. All at $\beta=2.2$.
  Pairs of lines are our $\pm 1\sigma$ estimates of the corresponding energies.}
\label{fig_EeffK2.2} 
\end{figure}

Of course, just as for our glueballs, this analysis depends on the
quality of our energy estimates. So in Fig.~\ref{fig_EeffK2.2}
we show the effective energy plots of the ground state flux tube
for every second value of $l$. Clearly the effective energy plateaux
are unambiguous (especially given the positivity constraint on $E_{eff}$)
except perhaps for $l=68$, which has a large error and so does not play
a significant role in our fits. In fact it is at small values of $l$
that the corrections to the linear behaviour become important, and
it is there that we have the best plateaux.

\subsection{`continuum' scaling}
\label{subsection_scaling}

\begin{figure}[htb]
\begin	{center}
\input	{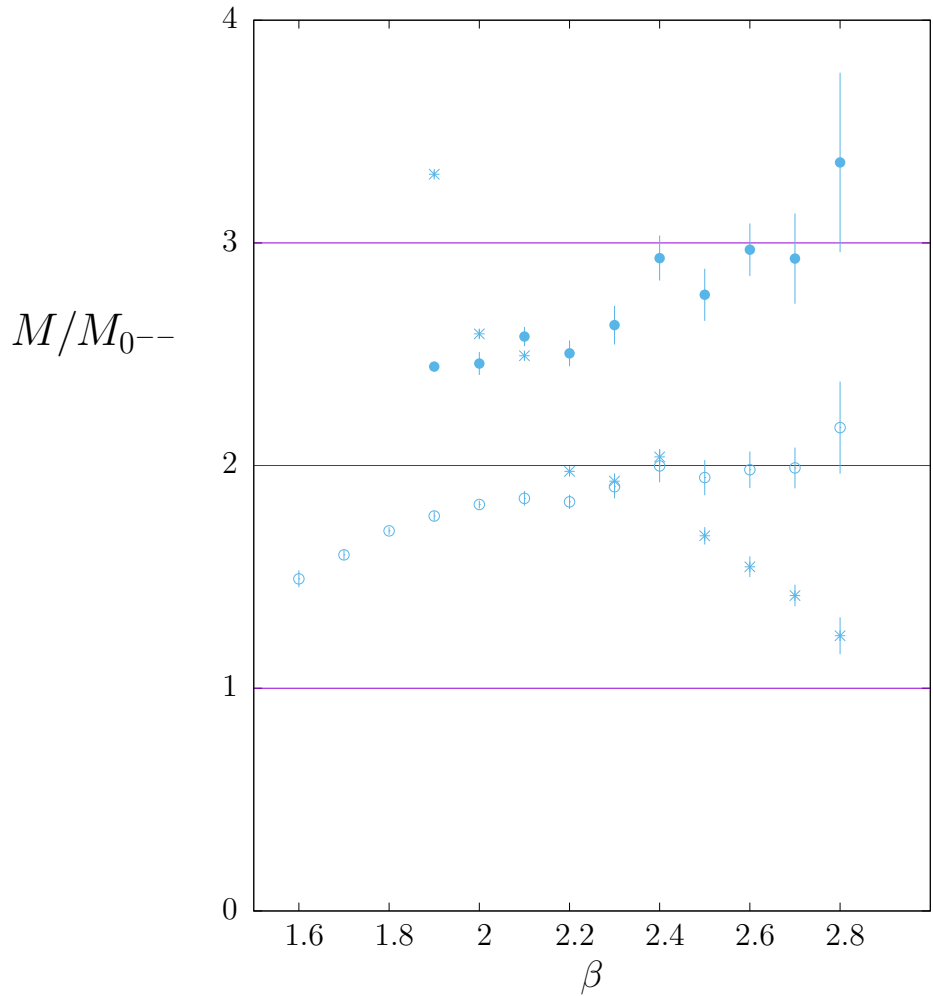}
\end	{center}
\caption{Masses of $0^{++}$ ground state ($\circ$)
  and $0^{--}$ first excited state ($\bullet$) in units of the mass of the
  $0^{--}$ ground state, versus $\beta$. Also $lM_{0^{--}}/3$ ($\star$), the
  rescaled spatial lattice size in units of the mass gap. }
\label{fig_G_b}
\end{figure}

In addition to the above detailed calculations at $\beta=2.2$ and 
$\beta=2.3$ we have also performed some calculations over the much
wider range $\beta\in [1.6,2.8]$. The parameters and results of these
calculations are listed in Table~\ref{table_datascan}. To keep the
calculation manageable we have not made the lattices large enough,
especially at large $\beta$, to be in the range of values of $lm_{0^{--}}$ where
our calculations at $\beta=2.2$ and $\beta=2.3$ told us that finite volume
corrections are negligible for all the tabulated quantities. So we can only be
confident in the values we obtain for the string tension and the
mass gap (except perhaps at $\beta=2.8$) since these quantities
are relatively insensitive to the volume, as we can see from
Tables~\ref{table_massV_b2.2} and \ref{table_massV_b2.3}. Nonetheless as
a test of how the finite volume corrections vary with $\beta$ it is interesting
to look at our results for the masses of the ground state $0^{++}$ and the
excited  $0^{--}$ rescaled by the mass gap. This we do in Fig.~\ref{fig_G_b},
where we also show how  the lattice size $lm_{0^{--}}$ varies with $\beta$ for our
lattices. What is interesting is that despite our decreasing volume (in physical
units) the ratios rapidly approach their free field values as we increase
$\beta$. This suggests that the  corrections that we have seen to free field
behaviour on small volumes are indeed due to the interactions arising
from higher order lattice spacing corrections, as we would expect.
Of course, if we were to decrease
the spatial size down towards $l\sim (1/T_c)$, then we would expect to see
a change. However this is not a concern for the continuum limit since
we have seen that $T_c/m_D \to \infty$ as $a \to 0$.

\begin{figure}[htb]
\begin	{center}
\input	{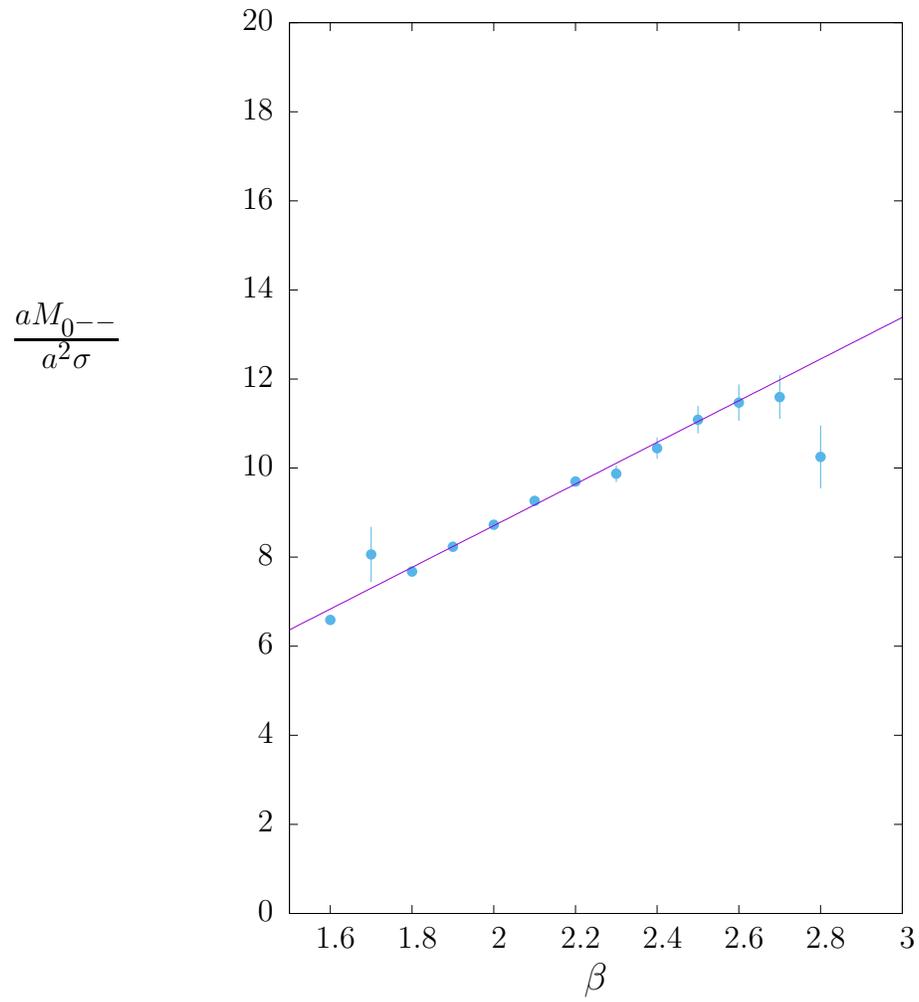}
\end	{center}
\caption{Mass of $0^{--}$ ground state divided by the string tension
  versus $\beta$.}
\label{fig_GK_b}
\end{figure}

Returning to the quantities we are confident in, i.e. the string tension
and the mass gap, we can ask if there is already evidence
for the asymptotic $M_{0{--}}/\sigma \propto \beta$ scaling implied
by eqns(\ref{eqn_mD},\ref{eqn_K}). We plot this ratio in Fig.~\ref{fig_GK_b}
and indeed we observe a very plausible linear growth with $\beta$
except for the extreme point at $\beta=2.8$. At this extreme value we begin
to see an indication  of a breakdown in the positivity of the flux loop
correlator which may be due to critical slowing down. In any case the
best linear fit shown in  Fig.~\ref{fig_GK_b} is 
\begin{equation}
  \frac{am_{0^{--}}}{a^2\sigma}  = -0.65(38) + 4.68(20)\beta  \qquad  1.7 \geq \beta \geq 2.7, \quad \frac{\chi^2}{n_{df}} = 0.72.
  \label{eqn_MKbfit}
\end{equation}
This fit is obtained by maximising the interval in $\beta$ over which we
can obtain a good linear fit.The evidence for the non-zero constant piece
in the fit is weak; indeed if restrict ourselves to the interval
$2.0 \geq \beta \geq 2.7$ we obtain $am_{0^{--}}/a^2\sigma = 0.01(66)+4.38(32)\beta$
for our best linear fit. In any case we note that the coefficient of $\beta$
in these fits is remarkably close to the classical value $\pi^2/2 \simeq 4.93$
that we infer from eqns(\ref{eqn_mD},\ref{eqn_K},\ref{eqn_KmDconst}).
  
\begin{figure}[htb]
\begin	{center}
\input	{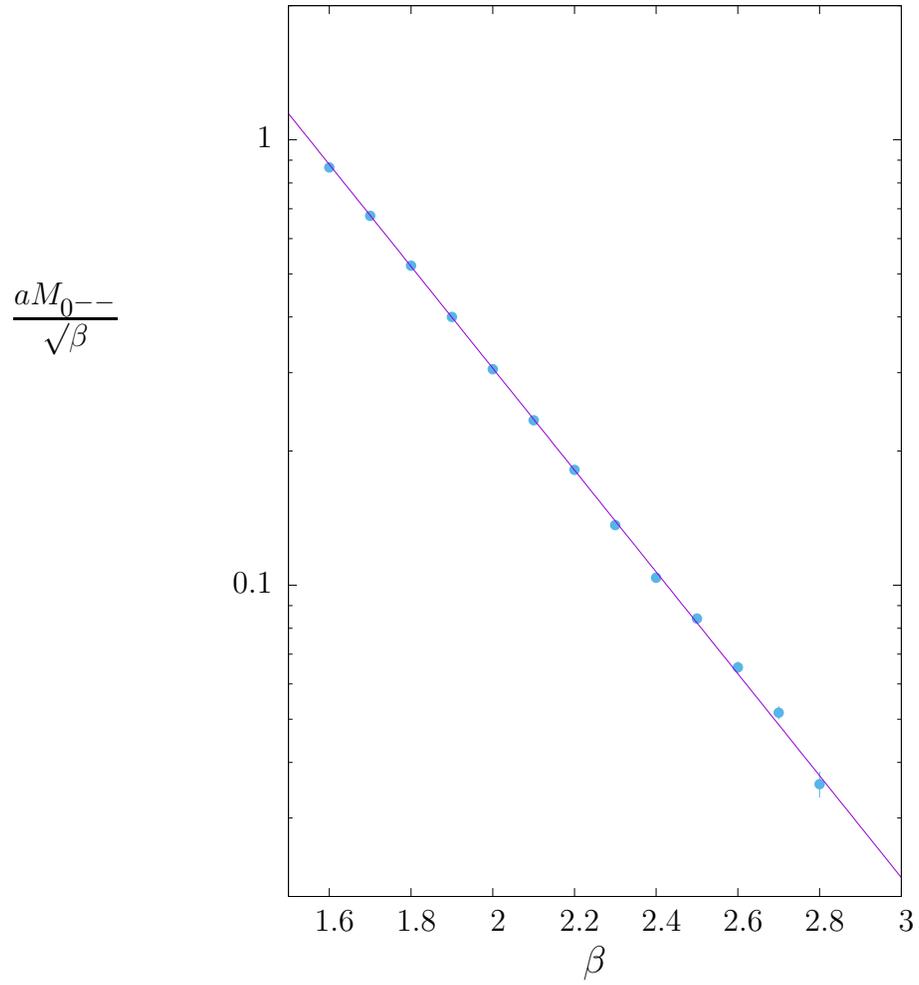}
\end	{center}
\caption{Mass of $0^{--}$ ground state rescaled by $\beta^{-1/2}$ versus $\beta$.}
\label{fig_m0mm_b}
\end{figure}

\begin{figure}[htb]
\begin	{center}
\input	{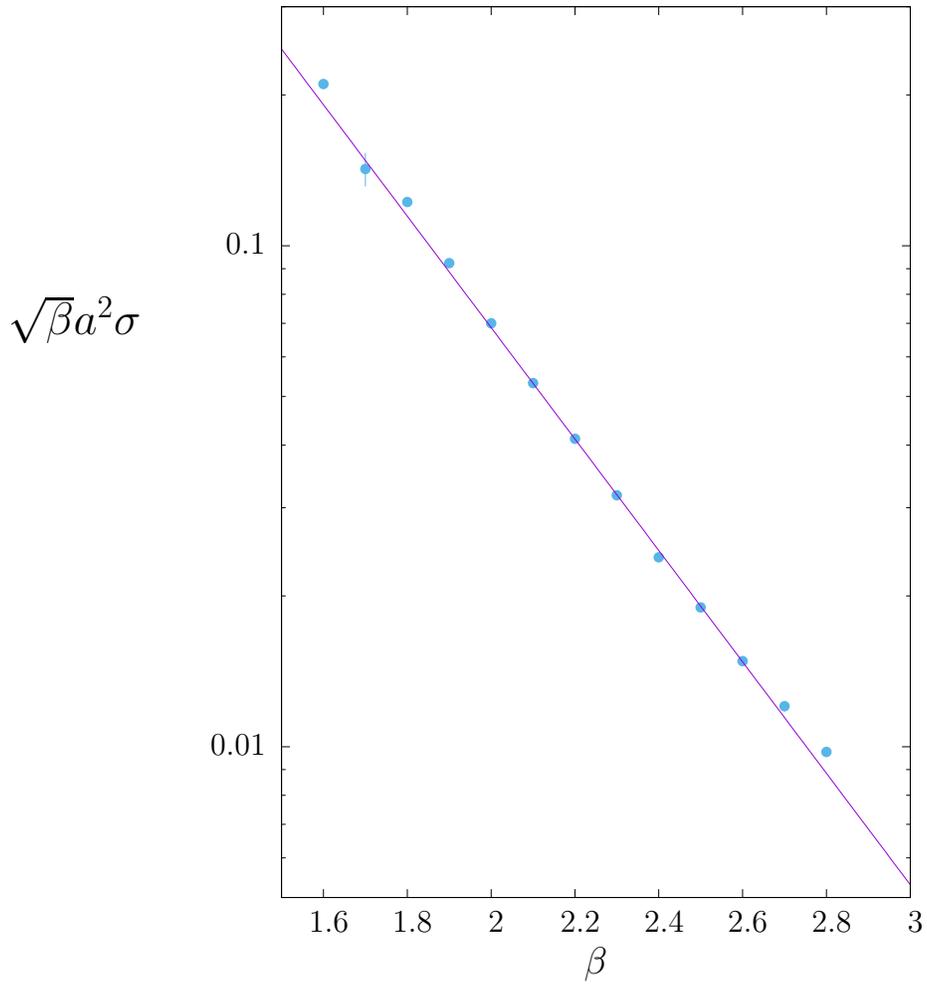}
\end	{center}
\caption{String tension rescaled by $\surd\beta$ versus $\beta$.}
\label{fig_K_b}
\end{figure}

It is also interesting to see whether the exponential behaviour of
$am_{0^{--}}\equiv am_D$ in eqn(\ref{eqn_mD}) is reproduced in our
calculations. We therefore plot in Fig.\ref{fig_m0mm_b} the quantity
$am_{0^{--}}/\surd\beta$ on a log-linear plot where an exponential appears
as a straight line. Clearly the data displays such a linear behaviour to
a good approximation and the fit shown is 
\begin{equation}
  \frac{am_{0^{--}}}{\surd\beta}  =  59.4(1.1) \exp\{-2.633(10)\beta\}  \qquad  1.7 \geq \beta \geq 2.8, \quad \frac{\chi^2}{n_{df}} = 1.34.
  \label{eqn_Mbfit}
\end{equation}
The theoretical classical value for the exponent is $0.2527\pi^2 \sim 2.494 \beta$
which is remarkably close to the value in our above fit. The coefficient of
the exponential is far from the classical value which is no surprise since it
will be sensitive to quantum fluctuations. We repeat the exercise for the
string tension, where we exclude the $\beta=2.8$ value for the same
reason as we did in eqn(\ref{eqn_MKbfit}). In this case the fit is poor unless
we also exclude several of the low $\beta$ values. The data is displayed in
Fig.\ref{fig_K_b} together with the fit
\begin{equation}
  \sqrt{\beta}a^2\sigma  =  11.5(6) \exp\{-2.561(22)\beta\}  \qquad  2.1 \geq \beta \geq 2.7, \quad \frac{\chi^2}{n_{df}} = 1.93.
  \label{eqn_Kbfit}
\end{equation}
where once again we observe that the exponent is remarkably close to the the
theoretical classical value. We note that, at least by eye, the fit 
appears to be reasonable over almost the whole range of $\beta$.
This is however something of an illusion due to the errors being mostly too
small to be visible on the plot. It may be puzzling that we can nonetheless
have an acceptable fit to  ${am_{0^{--}}}/{\surd\beta}$ over the range
$1.7 \geq \beta \geq 2.7$, but this is due to the fact that the errors
on that ratio are dominated by the errors from ${am_{0^{--}}}/{\surd\beta}$,
which are much larger than those on $a^2\sigma$.

We conclude that our
calculations show that both the mass gap and the string tension have
a functional dependence on $\beta$ that is very close to the theoretical
expectation, even in our modest range of $\beta$.

\section{Flux tubes: spectrum and massive modes}
\label{section_fluxtubes}

\begin{figure}[htb]
\begin	{center}
\input	{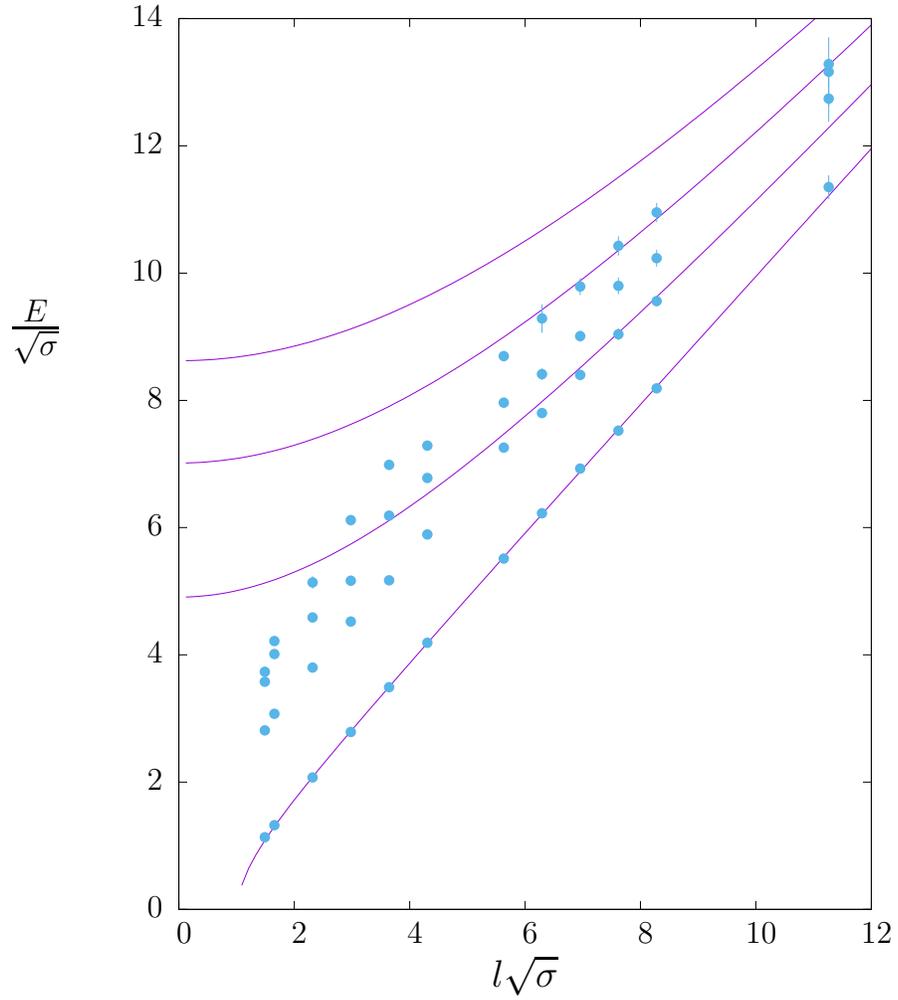}
\end	{center}
\caption{Positive parity flux tube spectrum versus length $l$, at $\beta=2.2$.
  Lines are naive string theory predictions.}
\label{fig_Ek1q0_b2.2}
\end{figure}

\begin{figure}[htb]
\begin	{center}
\input	{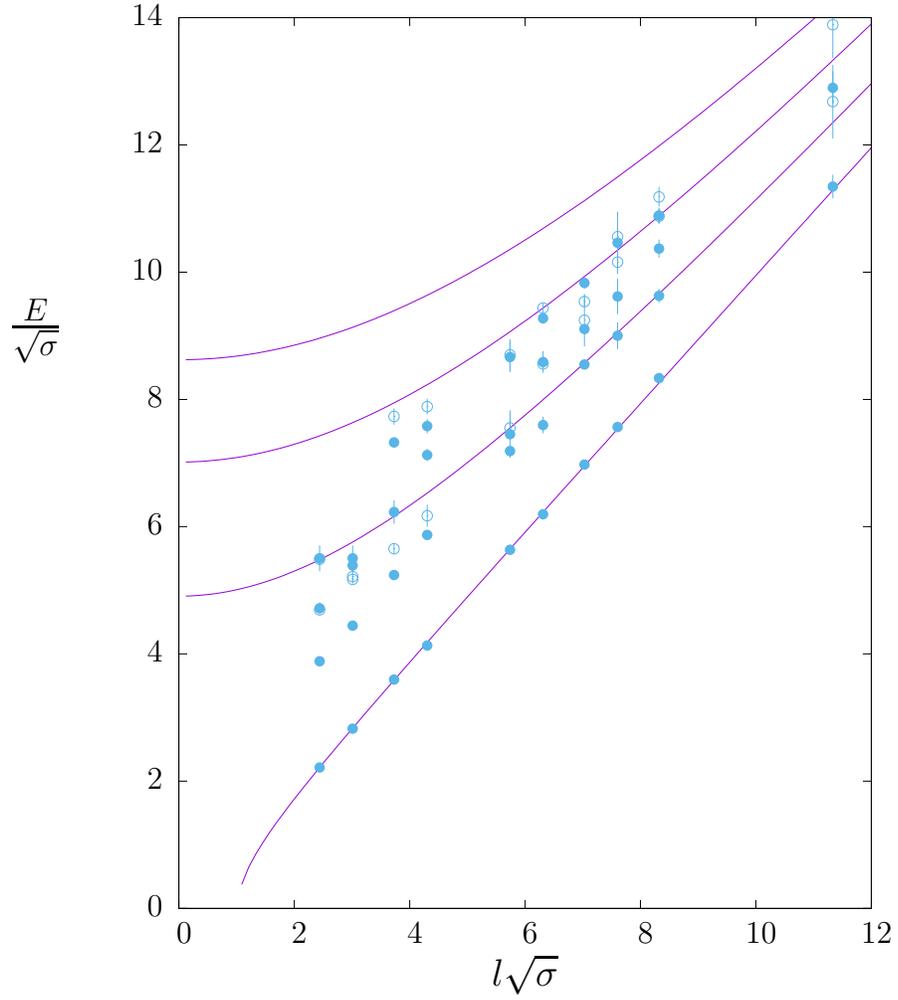}
\end	{center}
\caption{Flux tube spectrum versus length $l$, at $\beta=2.3$.
  States with positive ($\bullet$) and negative ($\circ$) parities. 
  Lines are naive string theory predictions.}
\label{fig_Ek1q0_b2.3}
\end{figure}

In calculating the spectrum of the winding flux tube we obtain not just
the energy of the ground state but also that of a number of excited states
of the flux tube.
In Tables~\ref{table_stringb2.2},\ref{table_stringb2.3} we present our
energy estimates for the lightest 4/5 states with positive parity, $P=+$,
at $\beta=2.2$, and in addition the lightest two $P=-$ states at $\beta=2.3$.
We plot the $\beta=2.2$ spectrum in Fig.~\ref{fig_Ek1q0_b2.2} and
the $\beta=2.3$ one in Fig.~\ref{fig_Ek1q0_b2.3}. The curves on the plot are
the predictions for the energies of the excited states as given by the string
theory calculations that predict eqn(\ref{eqn_NG}) for the ground state
\cite{AAMT_string1,AAMT_string2,AAMT_string3}.
These predictions actually work very well for the flux tube that carries
fundamental flux in $SU(N)$ gauge theories
\cite{AAMT_string1,AAMT_string2,AAMT_string3},
the more so as $N$ increases, and the theoretical reasons for this once
mysterious success are now understood
\cite{SD_string1,SD_string2,SD_string3}.
In the present case we see that there is no agreement except for the
ground state. In fact what we see here is much like the spectrum
of flux tubes in $SU(N)$ that carry flux in the $k=2A$ representation
(the totally antisymmetric piece of $f\otimes f$)
\cite{AAMT_string1,AAMT_string2,AAMT_string3}.
Just as in that case, we see a first excited state that has a
nearly constant gap from the ground state until it begins to meet
(and mix with) other excited states. This suggests that this state
is not a `stringy' excitation of the ground state, i.e. one which
contains massless transverse excitations on the world sheet, but is
simply the ground state with a massive excitation. If we plot the
gap of this state from the ground state as a function of $l$, we
see, in both Fig.~\ref{fig_DEexgs_b2.2} and Fig.~\ref{fig_DEexgs_b2.3},
that the associated  mass is consistent with that of the mass gap of the
bulk theory. This is in contrast to the $k=2A$ string in $SU(N)$
in $D=2+1$
\cite{AAMT_string1,AAMT_string2,AAMT_string3}
or the fundamental string in $D=3+1$  $SU(N)$
\cite{SD_mass,AAMT_string_D4_1,AAMT_string_D4_2}
where the mass of the excitation on the string world sheet is roughly
half the mass gap. Given that the $U(1)$ mass gap is just the screening
mass, which one can label as a `massive photon', one is tempted to
conjecture that the world sheet mass in these $SU(N)$ theories is that
of a `massive gluon'. The latter does not appear as an asymptotic state
in the bulk theory because of colour confinement, but in the $U(1)$ theory
the photon has no charge and so is not affected by confinement.

Of course this interpretation is speculative. An operator involving 
a single winding loop will have a non-zero overlap onto a state
consisting of a winding flux loop and a real massive photon 
(one will need some relative angular momentum to match the parities)
and perhaps this is what we are seeing. (However our experience
with $SU(2)$, where there is also no large-$N$ suppression,
suggests that the overlap should be small.) We also note that the
same mass appears in studies of the profile of the confining flux tube
\cite{Caselle_string1}
at large distances from the axis of the tube, as one would expect
since the asymptotic exponential decay of the profile should be
governed by the mass gap. Finally we recall that the very accurate
calculations in 
\cite{Caselle_string2}
for the ground state energy of flux tubes attached to static
sources provided evidence for the contribution of a rigidity term
in the world-sheet effective action, which involves a mass scale.
Our ground state energies are not nearly accurate enough to test
this idea for our closed flux tubes, so it would be very interesting
to know whether such a rigidity term can reproduce the interesting
behaviour of our first excited state.

\begin{figure}[htb]
\begin	{center}
\input	{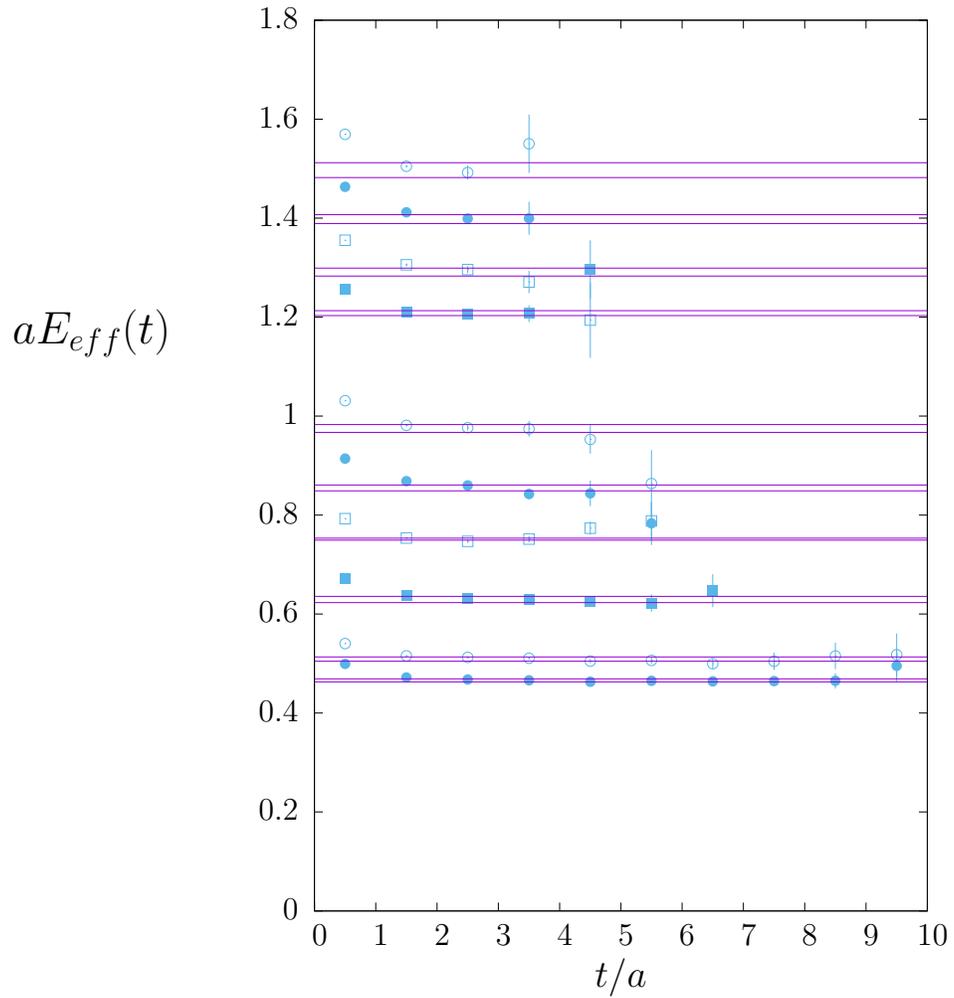}
\end	{center}
\caption{Effective energy of the first excited state of the winding flux tube
  with length $l=9,10,14,18,22,26,34,38,42,46$ in increasing order. At $\beta=2.2$.
  Pairs of lines are our $\pm 1\sigma$ estimates of the corresponding energies.}
\label{fig_EeffKex2.2} 
\end{figure}

This interference of a massive excitation on the world sheet depends
very much on our calculation of the excited flux tube energy being
reliable for the relevant range of $l$ in Figs~\ref{fig_DEexgs_b2.2}
and \ref{fig_DEexgs_b2.3}. So once again it is useful to display the
effective energy plots. This we do in Fig.~\ref{fig_EeffKex2.2} for
$\beta=2.2$ for all the values of $l$ that play a significant role here.
(At larger $l$ the levels meet and appear to repel.) The
identification of the effective energy plateaux appears reasonable,
especially at the small to medium values of $l$, which are the most relevant.

\section{Conclusions}
\label{section_conclusions}

After some preliminary calculations to localise the finite temperature
deconfining phase transition, so as to be confident that we would be
working in the low $T$ confined phase, we focused on the calculation
of the masses of `glueballs' of various $J^{PC}$ quantum numbers, and the
energies of confining flux tubes that wind around a periodic spatial
circle.

Our calculations of the lightest few glueball masses were entirely
consistent with the theoretical expectation that the theory becomes
a free field theory of massive particles as we approach the continuum
limit. Our detailed finite volume studies of the glueball spectrum,
at two modest values of $\beta=2/ag^2$  showed that the lightest
particle is indeed the $J^{PC}=0^{--}$ `massive photon'
whose inverse mass $1/m_D$ provides the screening length
of the monopole-instanton gas. We also saw that on
large volumes the mass of the lightest $0^{++}$ is approximately $2m_D$
and that of the first excited $0^{--}$ is approximately $3m_D $. This
is precisely what one expects if these states consist simply of the
minimum  number of the `massive photons' with the minimum momenta
consistent with those quantum numbers, and if there is no interaction
energy, i.e. the theory is a free field theory. Our scan over a wider
range of $\beta$ appeared to confirm that the finite volume effects
we saw were due to interactions mediated by lattice spacing corrections
arising from the plaquette action we use. More importantly the scan
provided a nice confirmation of the theoretically expected scaling
behaviour $am_D/a^2\sigma \propto \beta$ as well as the expected
exponential decay with $\beta$ of both $am_D/\beta^{1/2}$
and $\beta^{1/2}a^2\sigma$.

From our calculations of the ground state energy $E_f(l)$ of a winding flux
tube of length $l$, we obtained the string tension $\sigma$ using a standard
formula that encodes all the known universal corrections to the linear piece.
This formula turns out to fit our energies very accurately down to small
values of $l$, albeit not as small as in $SU(N)$ gauge theories.
We also calculated the energies of the lightest few excited states of the
flux tube. These show little sign of a simple `stringy' behaviour,
and this is reminiscent of what one sees in $SU(N)$
for flux tubes that carry a flux in
a representation other than the fundamental -- in particular the $k=2A$
which is the antisymmetric piece of $f\otimes f$. Just as in the latter
case the first excited state is consistent with being the ground state
with a single massive excitation. The mass of this excitation is simply
$\simeq m_D$. This differs from what we see in $SU(N)$ in $D=3+1$ where
the mass of the world-sheet excitation is about one-half of the mass gap
of the bulk theory. This may motivate the speculation that the latter mass
may be that of a confined `massive gluon' just like the non-confined
`massive photon' on the $U(1)$ world sheet.

\section*{Acknowledgements}
AA has been financially supported by VI-SEEM and OPEN SESAME Horizon 2020 projects. 
MT acknowledges support by All Souls College and Oxford Theoretical Physics.
The numerical computations were carried out on the computing cluster
in Oxford Theoretical Physics.

\clearpage

\begin{table}[ht]
\begin{center}
  \begin{tabular}{|c|ccc|}
    \hline
    $lattice$ & $aM_{0^{--},gs}$ & $aM_{0^{--},ex1}$ &  $aM_{0^{++},gs}$ \\  \hline
 $68^236$    & 0.2670(13)  & 0.7709(38)  & 0.5376(29)     \\
 $50^236$    & 0.2676(21)  & 0.7728(38)  & 0.5401(35)     \\
 $42^236$    & 0.2670(23)  & 0.7662(70)  & 0.5295(61)     \\
 $34^236$    & 0.2671(27)  & 0.745(12)   & 0.5316(68)     \\
 $26^236$    & 0.2636(22)  & 0.705(17)   & 0.5040(65)     \\
 $22^236$    & 0.2624(30)  & 0.690(12)   & 0.4995(66)     \\
 $18^236$    & 0.2647(29)  & 0.690(16)   & 0.4544(51)     \\
 $14^236$    & 0.2816(40)  & 0.76(11)    & 0.3912(25)     \\ \hline
 \end{tabular}
  \caption{Masses of the lightest and first excited $J^{PC}=0^{--}$ glueballs, and
    of the lightest $0^{++}$ glueball, on various spatial volumes at $\beta=2.2$.}
\label{table_massV_b2.2}
\end{center}
\end{table}

\begin{table}[ht]
\begin{center}
  \begin{tabular}{|c|ccc|}
    \hline
    $lattice$ & $aM_{0^{--},gs}$ & $aM_{0^{--},ex1}$ &  $aM_{0^{++},gs}$ \\  \hline
 $60^260$ & 0.2111(22)    &  0.6118(68)  &  0.4289(38)  \\ 
 $54^260$ & 0.2102(22)    &  0.6244(53)  &  0.4247(42)  \\ 
 $36^248$ & 0.2059(24)    &  0.5954(82)  &  0.4168(47)  \\ 
 $30^248$ & 0.2107(28)    &  0.5740(83)  &  0.3972(50)  \\ 
 $24^232$ & 0.2138(38)    &  0.5492(88)  &  0.3924(40)  \\ \hline
 \end{tabular}
  \caption{Masses of the lightest and first excited $J^{PC}=0^{--}$ glueballs, and
    of the lightest $0^{++}$ glueball, on various spatial volumes at $\beta=2.3$.}
\label{table_massV_b2.3}
\end{center}
\end{table}

\begin{table}[ht]
\begin{center}
  \begin{tabular}{|c|c|ccccc|}
    \hline
 $l$ & $l_\perp$ & $aE^{P=+}_{gs}$ & $aE^{P=+}_{ex1}$  & $aE^{P=+}_{ex2}$ & $aE^{P=+}_{ex3}$& $aE^{P=+}_{ex4}$ \\ \hline
 9   &  75   & 0.1878(11)  & 0.4659(31)  & 0.5926(51) & 0.6184(53)   &   \\
 10  &  50   & 0.2190(12)  & 0.5090(44)  & 0.6647(42) & 0.6984(60)   &   \\
 14  &  64   & 0.3434(25)  & 0.6295(53)  & 0.760(11)  & 0.851(16)    &   \\
 18  &  50,64,90   & 0.4617(17)  & 0.7492(30)  & 0.8555(50) &  1.013(53)  &   \\
 22  &  44,64      & 0.5785(25)  & 0.8565(50)  & 1.025(10)  &  1.157(7)   & 1.15(4)  \\
 26  &  42,64      & 0.6941(24)  & 0.9758(65)  & 1.123(9)   &  1.207(11)  & 1.20(4)  \\
 34  &  34   & 0.9130(47)  & 1.202(6)  & 1.319(13)  & 1.440(12) & 1.51(4)  \\
 38  &  38   & 1.0310(36)  & 1.259(26) & 1.393(15)  & 1.538(37) & 1.59(3)  \\
 42  &  42   & 1.1472(57)  & 1.391(14) & 1.492(14)  & 1.621(22) & 1.62(4)  \\
 46  &  46   & 1.246(10)   & 1.497(15) & 1.623(22)  & 1.727(25) & 1.69(4)  \\
 50  &  50   & 1.356(9)    & 1.583(13) & 1.695(22)  & 1.814(25) & 1.85(5)  \\
 68  &  68   & 1.880(31)   & 2.11(6)   & 2.18(9)    & 2.20(12)  &   \\  \hline
\end{tabular}
  \caption{Energies of the lightest states of a flux tube of length $l$
    and positive parity $P$, winding around the $x$ direction of the $l\times l_\perp$
    spatial volume at $\beta=2.2$.}
\label{table_stringb2.2}
\end{center}
\end{table}

\begin{table}[ht]
\begin{center}
  \begin{tabular}{|c|c|cccccc|}
    \hline
    $l$ & $l_\perp$ & $aE^{P=+}_{gs}$ & $aE^{P=+}_{ex1}$  & $aE^{P=+}_{ex2}$ & $aE^{P=+}_{ex3}$ & $aE^{P=-}_{gs}$ & $aE^{P=-}_{ex1}$ \\ \hline
 17   & 74 & 0.3179(12)   & 0.5571(47)  &  0.677(14)  &  0.789(30)  & 0.673(7)    & 0.78(14)  \\
 21   & 74 & 0.4054(16)   & 0.6377(65)  &  0.773(12)  &  0.881(22)  & 0.742(7)    & 0.748(16)  \\
 26   & 74 & 0.5159(21)   & 0.7519(66)  &  0.894(27)  &  1.051(12)  & 0.811(13)   & 1.109(19)  \\
 30   & 50 & 0.5928(46)   & 0.842(10)   &  1.022(14)  &  1.088(17)  & 0.886(25)   & 1.132(18)  \\
 40   & 40 & 0.8085(34)   & 1.031(10)   &  1.069(54)  &  1.243(34)  & 1.083(18)   & 1.248(35)  \\
 44   & 44 & 0.8886(47)   & 1.090(19)   &  1.232(25)  &  1.331(12)  & 1.228(8)    & 1.354(11)  \\
 49   & 49 & 1.0009(54)   & 1.227(8)    &  1.306(11)  &  1.410(14)  & 1.368(13)   & 1.326(59)  \\
 53   & 53 & 1.0853(54)   & 1.291(30)   &  1.380(40)  &  1.501(70)  & 1.457(15)   & 1.515(21)  \\
 58   & 58 & 1.1957(86)   & 1.381(16)   &  1.488(21)  &  1.562(18)  & 1.561(17)   & 1.604(22)  \\
 79   & 79 & 1.628(27)    & 1.850(38)   &             &             &             &            \\ \hline
\end{tabular}
  \caption{Energies of the lightest states of a flux tube of length $l$ winding around the $x$
    direction of the  $l\times l_\perp$ spatial volume at $\beta=2.3$, for both parities $P$.}
\label{table_stringb2.3}
\end{center}
\end{table}

\begin{table}[ht]
\begin{center}
  \begin{tabular}{|cc|cccc|}
    \hline
 $\beta$ & $lattice$ & $aM_{0^{--},gs}$ & $aM_{0^{--},ex1}$ &  $aM_{0^{++},gs}$ & $a\surd\sigma$  \\  \hline
 2.8  & $62^2110$  & 0.0598(40)   &  0.201(20)  & 0.1298(88)   &  0.07638(64)   \\ 
 2.7  & $50^296$   & 0.0850(29)   &  0.249(15)  & 0.1691(52)   &  0.08562(106)   \\ 
 2.6  & $44^296$   & 0.1054(32)   &  0.313(9)   & 0.2088(59)   &  0.09586(88)   \\ 
 2.5  & $38^296$   & 0.1330(31)   &  0.368(13)  & 0.2588(86)   &  0.10953(84)   \\ 
 2.4  & $38^264$   & 0.1610(28)   &  0.472(14)  & 0.3218(104)  &  0.12415(91)   \\ 
 2.3  & $28^264$   & 0.2068(37)   &  0.544(15)  & 0.3940(83)   &  0.14473(43)   \\ 
 2.2  & $22^248$   & 0.2692(27)   &  0.674(14)  & 0.4945(71)   &  0.16661(30)   \\ 
 2.1  & $22^248$   & 0.3400(27)   &  0.877(13)  & 0.630(10)    &  0.19159(46)   \\ 
 2.0  & $18^240$   & 0.4320(31)   &  1.062(21)  & 0.7884(50)   &  0.22250(39)   \\ 
 1.9  & $18^240$   & 0.5514(26)   &  1.348(8)   & 0.978(11)    &  0.25877(28)   \\ 
 1.8  & $18^240$   & 0.6996(21)   &             & 1.194(11)    &  0.3019(18)   \\ 
 1.7  & $14^232$   & 0.8799(24)   &             & 1.407(19)    &  0.330(13)   \\ 
 1.6  & $14^232$   & 1.0962(31)   &             & 1.635(41)    &  0.4078(16)   \\  \hline
\end{tabular}
  \caption{Energies of the lightest glueball states and the string tension
    at various values of $\beta$ on the lattices shown.}
  \label{table_datascan}
\end{center}
\end{table}

\clearpage

\begin{figure}[htb]
\begin	{center}
\input	{plot_VGb2.2.tex}
\end	{center}
\caption{$0^{--}$ ground state ($\bullet$), $0^{++}$ ground state ($\circ$)
  and $0^{--}$ first excited state ($\blacklozenge$) on spatial volumes $V=l^2$
  at $\beta=2.2$. Lines are once, twice, and thrice the large volume $0^{--}$
  ground state mass. }
\label{fig_VGb2.2}
\end{figure}

\begin{figure}[htb]
\begin	{center}
\input	{plot_VGb2.3.tex}
\end	{center}
\caption{$0^{--}$ ground state ($\bullet$), $0^{++}$ ground state ($\circ$)
  and $0^{--}$ first excited state ($\blacklozenge$) on spatial volumes $V=l^2$
  at $\beta=2.3$. Lines are once, twice and thrice the large volume $0^{--}$
  ground state mass. }
\label{fig_VGb2.3}
\end{figure}

\begin{figure}[htb]
\begin	{center}
\input	{plot_DEexgs_b2.2.tex}
\end	{center}
\caption{Difference between ground state energy and first ($\bullet$),
  second ($\circ$) and third ($\Box$) excitations in Fig.~\ref{fig_Ek1q0_b2.2}.
  Horizontal line is the bulk mass gap. Falling lines are differences between
  ground and first/second excited string theory energies.}
\label{fig_DEexgs_b2.2}
\end{figure}

\begin{figure}[htb]
\begin	{center}
\input	{plot_DEexgs_b2.3.tex}
\end	{center}
\caption{Difference between ground state energy and first ($\bullet$),
  second ($\circ$) and third ($\Box$) $P=+$ excitations in Fig.~\ref{fig_Ek1q0_b2.3}.
  Horizontal line is the bulk mass gap. Falling lines are differences between
  ground and first/second excited string theory energies.}
\label{fig_DEexgs_b2.3}
\end{figure}

\end{document}